\documentclass[aps,showpacs]{revtex4}
\usepackage{graphicx}
\usepackage[all]{xy}
\usepackage{amsmath}
\usepackage{amssymb}
\newcommand{\be}{\begin{equation}}
\newcommand{\ee}{\end{equation}}
\newcommand{\ben}{\begin{eqnarray}}
\newcommand{\een}{\end{eqnarray}}
\newcommand{\bes}{\begin{subequations}}
\newcommand{\ees}{\end{subequations}}
\newcommand{\wt}{\widetilde}

\newcommand{\bb}{\bibitem}
\newcommand{\sech}{{\rm sech}}

\begin{document}

\title{Classical behavior of deformed sine-Gordon models}
\author{D. Bazeia,$^1$ L. Losano,$^1$ J.M.C. Malbouisson,$^2$ and R. Menezes$\,^1$}
\affiliation{$^1$ Departamento de F\'{\i}sica, Universidade Federal
da Para\'{\i}ba, 58051-970 Jo\~ao Pessoa, PB, Brazil\\
$^2$ Instituto de F\'{\i}sica, Universidade Federal da Bahia, 40210-340 Salvador, BA, Brazil}
\date{\today}
\begin{abstract}
In this work we deform the $\phi^4$ model with distinct deformation functions, to propose a diversity of sine-Gordon-like models.
We investigate the proposed models and we obtain all the topological solutions they engender. In particular,
we introduce non-polynomial potentials which support some exotic two-kink solutions. 
\end{abstract}
\pacs{11.27.+d, 11.25.-w, 82.35.Cd}
\maketitle

\section{Introduction}
\label{s1}

The sine-Gordon equation \cite{sg1,sg2,sg3,sg4,sg5} has been used in many different areas of applications in nonlinear science \cite{b1,mol,b2,no,vi,fw,b3,ms,dna}.
Among the several possibilities of study, the specific investigations on classical, semiclassical and other aspects of sine-Gordon, double sine-Gordon and related models \cite{g00,g01,g02,g03,g1,g2,g3,g4,g5,g6,g7,g8,g9,g10,g11,g13,g14,g15,g16,g17,g18,g19,g20,g21,g22} have motivated us to propose the present study.
 
An interesting variation of the sine-Gordon model is the double sine-Gordon model, which has attracted recent attention as a non-integrable model that can be studied by the approach used in the context of integrable models. We believe that the new models here introduced may perhaps be used to such study, in particular in the lines of \cite{g2,g7,g14,g19} to check if the semiclassical study goes as in the double sine-Gordon case. Another line of study could be the search of internal modes \cite{g1,g4,g6}, which seems to appear after deforming an integrable model \cite{g1}. The sine-Gordon model has also been of interest for String Theory, since it seems to be related to the classical string on specific manifolds \cite{st1,st2}.

Phenomenologically, the sine-Gordon model has been used in a variety of situations, in molecular \cite{mol} and DNA \cite{dna} dynamics, in ferromagnetic
waves \cite{fw}, nonlinear optics \cite{no} and in several other areas \cite{vi,ms}. The sine-Gordon model that we investigate are modifications of the original model, and they may also be used in application since they are in general richer then the standard model, depending on new parameters which control important features of the potential, leaving extra room to fuel applicability.

The sine-Gordon-like models that we investigate in the present work are all obtained as deformation of the $\phi^4$ model, so we start this work in Sec.~{\ref{s2}}, where we review some basic facts on the deformation procedure first introduced in \cite{dd1} and further explored in Refs.~{\cite{dd2,dd3,dd4,dd5,du1,du2,dd6}. In Sec.~{\ref{s3}} we introduce and study the defect solutions for several distinct families of
sine-Gordon-like models, and in Sec.~{\ref{s4}} we present our ending comments.

\section{Generalities}
\label{s2}

We start considering a model described by a single real scalar field $\phi$, in $(1,1)$ space-time dimensions, with Lagrange density
\begin{equation}\label{sm} {\cal
L}=\frac12\partial_\mu\phi\,\partial^\mu\phi-V(\phi) .
\end{equation}
We use the metric $(+,-)$ and work with dimensionless fields and coordinates, for simplicity. The equation of motion for $\phi=\phi(x,t)$ is
\be\label{em}
\partial_\mu\partial^\mu\phi+V^\prime(\phi)=0
\ee
where the prime stands for derivative with respect to the argument. For a static field $\phi=\phi(x)$ we have
\begin{equation}\label{em1}
\frac{d^2\phi}{dx^2}=V^{\prime}(\phi ) ,
\end{equation}
This equation has a first integral given by
\begin{equation}\label{em1phi2}
\left(\frac{d\phi}{dx}\right)^2 = 2\, V(\phi) + c ,
\end{equation}
where $c$ is an integration constant, which can be identified with the stress of the static solution \cite{blmo}. 

The energy associated to a static solution is
\begin{equation}\label{H}
E=\frac12 \int_{-\infty}^{+\infty} dx \left( \left(\frac{d\phi}{dx}\right)^2 + 2\,V(\phi) \right) .
\end{equation}
Thus, $c\neq 0$ in Eq.~(\ref{em1phi2}) corresponds to infinite energy solutions. However, if the potential has at least one
critical point at $\bar\phi$, that is $V^{\prime}(\bar\phi)=0$, for which $V(\bar\phi)=0$, any solution of (\ref{em1phi2}) satisfying the boundary conditions
\begin{equation}\label{BC}
 \phi(x\to-\infty)\to\bar\phi \,\, ,
\,\,\,\,\, \frac{d\phi}{dx}(x\to-\infty)\to 0 \,,
\end{equation}
will require that $c=0$, which provides a necessary condition for the presence of finite energy solutions. Such a circumstance is not sufficient to ensure the existence of finite energy solution as shown, for example, by the static solutions $\phi_\pm(x) = a e^{\pm x}$ ($a$ constant) of the free Klein-Gordon
model, where $V(\phi)=({1}/{2})\phi^2$. 

If we write $c=0$ and $V=(1/2)(dW/d\phi)^2=W_\phi^2/2$ in \eqref{em1phi2}, we can get to
\be\label{foeq}
\frac{d\phi}{dx}=W_\phi
\ee 
where the $\pm$ has been absorbed into $W.$ The energy of the solutions of the above first-order equation \eqref{foeq} can be written as $E=|\triangle W|,$ where
$\triangle W=W(\phi(x\to\infty))-W(\phi(x\to-\infty)),$ and this very much eases the investigation \cite{b}. 

We investigate linear stability of the static solution using $\phi(x,t)=\phi(x)+\eta(x,t)$ into the equation (\ref{em}) to get, for $\eta(x,t)=\sum_n\eta_n(x)\cos(w_nt),$ the Schr\"odinger-like equation $H\eta_n(x)=w_n^2\eta_n(x),$ with Hamiltonian
\be\label{schro}
H=-\frac{d^2}{dx^2}+V^{\prime\prime}(\phi)
\ee
where $V^{\prime\prime}(\phi)$ has to be calculated at the static solution $\phi=\phi(x),$ to become a function of $x:$ $V^{\prime\prime}(\phi)\equiv U(x).$

There are two very well known models described by real scalar field, which are described by 
\be\label{potphi4}
V(\phi)=\frac12(1-\phi^2)^2
\ee
and by
\be\label{sg}
V(\phi)=\frac12 \cos^2(\phi)
\ee
The first is the $\phi^4$ model, and has kink-like solutions given by $\phi_\pm(x)=\pm\tanh(x);$ the second is the sine-Gordon model, and it has solutions $\phi_\pm(x)=\pm\arcsin(\tanh(x))+k\pi,$ where $k=0,\pm1,\pm2,...$ specifies one among the infinity of sectors of the model. The study of stability leads to the following potentials: for the $\phi^4$ model
\be
U(x)=4-6\,{\rm sech}^2(x)
\ee
and for the sine-Gordon
\be\label{sgq}
U(x)=1-2\,{\rm sech}^2(x)
\ee
In the first case, the potential of the Schr\"odinger-like equation supports two bound states, the massless zero mode and a massive eigenstate. In the second case
there is only one bound state, the massless zero mode. By the way, it is not hard to see that the zero mode is proportional to the derivative of the static solution, a fact which follows almost directly from the Sch\"odinger-like equation with Hamiltonian \eqref{schro}. The specific features of these Schr\"odinger-like problems have been recently explored in Refs.~{\cite{g10,g13}}, to reconstruct the corresponding field theory model. 

When we write the potential in the form $V=W^2_\phi/2$ we have that
\be
V^{\prime\prime}=W^2_{\phi\phi}+W_\phi W_{\phi\phi\phi}
\ee
in a way such that the Hamiltonian \eqref{schro} now can be written as $H=S^\dag S,$ for $S$ being the first-order operator
\be
S=-\frac{d}{dx}+W_{\phi\phi}
\ee
which shows that the solutions of the first-order equation \eqref{foeq} are linearly stable.

The deformation method, introduced in Ref.~{\cite{dd1}} and extended in Refs.~{\cite{dd2,dd3,dd4}}, consists in the following general
prescription. Denote the deformed field by $\chi$ and assume that it is mapped into the starting field through the deformation function $f$, that is,
\begin{equation}\label{phichi}
\phi = f(\chi), \,\, \,\,\,\,\, \chi = f^{-1}(\phi) .
\end{equation}
If $\phi(x)$ is a static solution of the starting model (\ref{sm}), then $\chi(x)=f^{-1}(\phi(x))$ is a static solution of the deformed model,
described by the Lagrange density
\begin{equation}\label{m2}
{\wt{\cal L}}=\frac12\partial_\mu\chi\partial^\mu\chi-{\wt V}(\chi) ,
\end{equation}
where the deformed potential ${\wt V}(\chi)$ has the form
\begin{equation}\label{vd1}
{\wt V}(\chi)=\frac12\frac{2\,V(f(\chi))+c}{[df/d\chi]^2},
\end{equation}
with $c$ being the constant appearing in Eq.~(\ref{em1phi2}), associated with the solution $\phi(x)$ of the starting problem. In this case, the
potential of the Schr\"odinger-like equation appearing in the study of the stability of the deformed classical solution has the form
\be
{\wt U}(x)=\left.\frac{d^2{\wt V}(\chi)}{d\chi^2}\right|_{\chi=\chi(x)},
\ee
and can be directly obtained from the deformation prescription.

The deformation method is powerful and interesting. Mathematically, it provides a way to construct an uncountable number of new nonlinear
differential equations of the type of Eq.~(\ref{em1}), and their solutions, starting from any known solvable model. Physically, it
allows to deform topological and non-topological defects, controlling their energy and width, and generating a large amount of
new defect solutions having distinct characteristics from the original ones.

The method has been used in a broad sense. In Ref.~\cite{dd1}, we considered bijective deformation functions, which do not alter the
topological nature of the defects, but permits the control of their energy and width. Then, considering non-bijective deformations
{\cite{dd2}}, with $f^{-1}$ being a multivalued function, we show how to deform a non-topological (lump-like) defect into a kink-like
(topological) one. Functions $f$, for which the equation $f^{\prime}(\phi)=0$ has finite solutions, can be also considered if
the starting potential is such that ${\wt V}(\chi)$ does not possess singularities. In all these cases, the analysis was restricted to
models having finite energy solutions. A further development {\cite{dd3}} extended the method, presenting examples where infinite
energy solutions, with $c\neq 0$ (as those of the free Klein-Gordon model), can be deformed into finite energy ones, thus having
physical significance, using deformation functions which are real valued only when defined in a finite interval of {\bf R}. We have also
used interesting deformation functions, from which we have generated distinct families of polynomial potentials \cite{dd4}.
Other investigations have dealt with more general issues, extending the deformation procedure to tachyonic models \cite{dd5}
and to models described by two scalar fields \cite{dd6}. 

The deformation method can be further extended to generate acceptable solutions starting from singular solutions of known
models. In the next section, we focus on the important, physically relevant,
models of the sine-Gordon type. Before doing this, however, let us exemplify the deformation procedure with the two models given above. 
We consider the model (\ref{potphi4}) and the deformation functions $f_\pm(\chi)=\pm\sin(\chi).$ This function and the prescription
(\ref{vd1}), with $c=0,$ leads immediately to the sine-Gordon potential (\ref{sg}) and its solutions $\chi(x) = \pm\arcsin[\tanh(x)] + k \pi$,
where $k$ is an integer that specifies the branches of the inverse $f^{-1}_k(\phi)=\pm\arcsin(\phi)+k\pi$ (from here on,
$\arcsin(\phi)$ denotes the principal branch with values in $\left[-{\pi}/{2} , {\pi}/{2} \right]$ and, similarly, for any
other multivalued inverse function). While the kink solutions of the $\phi^4$ model connect the minima $+1$ to $-1,$ the infinitely many
topological solutions of the sine-Gordon model run between adjacent minima $\pi(k-1/2)$ and $\pi(k+1/2)$, as illustrated in Fig.~1.

\begin{figure}[ht]
\includegraphics[width=16cm]{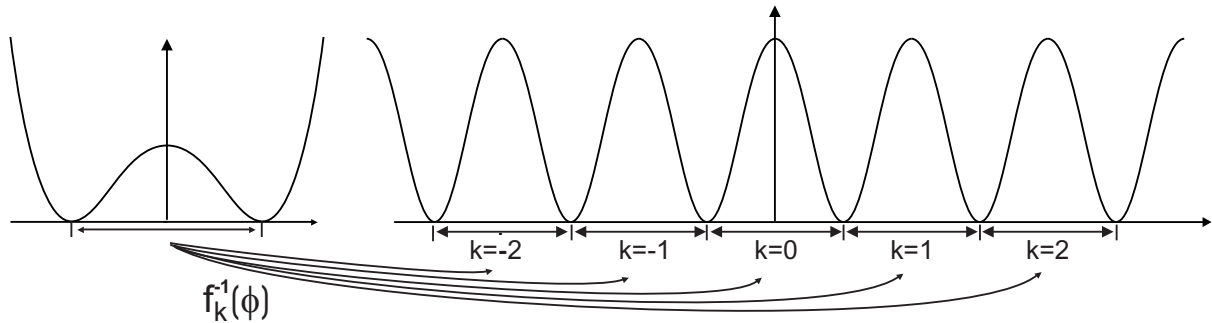}
\caption{The potentials of the $\phi^4$ \eqref{potphi4} (left) and sine-Gordon \eqref{sg} (right) models, and the mapping of the topological
sector of the left model into the topological sectors of the right model through the branches of the inverse of the deformation function $f^{-1}_k(\phi)$.}\label{phi4sine}
\end{figure}

\section{Some new sine-Gordon models}
\label{s3}

In this section, we explore new possibilities, considering three distinct classes of deformation functions to obtain new sine-Gordon-like models that support topological solutions. All the deformed models are obtained via specific deformations of the $\phi^4$ model, described by the potential (\ref{potphi4}).
In this case, since the $\phi^4$ model is polynomial, to get to sine-Gordon-like models we need to use periodic functions as the deformation functions.
We do this below, where we investigate explicit models.

\subsection{Type-I models}

We turn our attention to a class of models which are obtained from the $\phi^4$ model (\ref{potphi4})
using the deformation functions
\begin{equation}
f_n(\chi)=1-2\sin^n(\chi).
\end{equation}
for $n$ being a constant parameter which specify the deformation function. These functions are acceptable functions if $n$ obeys $\sin^{n}(\chi)\in {\bf R}$ for all $\chi(x)\in{\bf R}$. This imposes that $n$ may be a even integer or rational even/odd number.
In general, the deformed potentials are given by
\begin{equation}\label{potdsg1}
{\wt V}(\chi)=\frac2{n^2}\tan^2(\chi)\left(1-\sin^n(\chi)\right)^2 ,
\end{equation}
where for $n=2$ we have the usual sine-Gordon case.
These potentials have an infinite set of minima given by
\be
\bar\chi=\frac{k}2\pi ,
\ee
where $k$ is an integer. There are two kind of minima: for $k=2l$, even, at the points ${\bar\chi}_e=l\pi$ we have
$d^2{\wt V}/d\chi^2=4/n^2$ and so the concavity depends on the parameter $n$ for these minima; on the other hand, for odd $k$, we
find that the concavity is the same at the points ${\bar\chi}_o=(k+1/2)\pi$ where $d^2{\wt V}/d\chi^2=1.$ The maxima of the potentials obey the following relation
\be 
\sin({\bar\chi})^{2+n}-\frac{n+1}{n}\sin({\bar\chi})^n+\frac1{n}=0.
\ee

Here we notice that the potential (\ref{potdsg1}) can be written as in the form ${\wt V}(\chi)=(1/2)(W_{\chi})^2$, where $W_{\chi}=dW/d\chi=(2/n)\tan(\chi)(1-\sin^n(\chi))$. For $n$ even and positive, the superpotential functions $W(\chi)$ can be written as
\be
W_n(\chi)=\frac1n\sum_{j=1}^{n/2} \frac{1}{j} \sin^{2j}(\chi) .
\ee
From this result, we can find the energy of the topological solutions that connect two adjacent minima of the potential, $E_n = W_n(\chi_b) -W_n(\chi_a)$. Since $W(\chi_a)=0$ and $W(\chi_b)=\Psi(n/2+1)+\gamma,$ where $\Psi(z)$ and $\gamma$ are the digamma function and the Euler constant respectively, we find
\be\label{eny1}
E_n=\frac1n\left[\Psi(n/2+1)+\gamma\right].
\ee
In the limit $n\to 0$ we obtain $E_0=\pi^2/12$, and we have the values $E_2=1/2$, $E_4=3/8$, and $E_6=11/36$.

\begin{figure}[ht!]
\includegraphics[{width=8cm}]{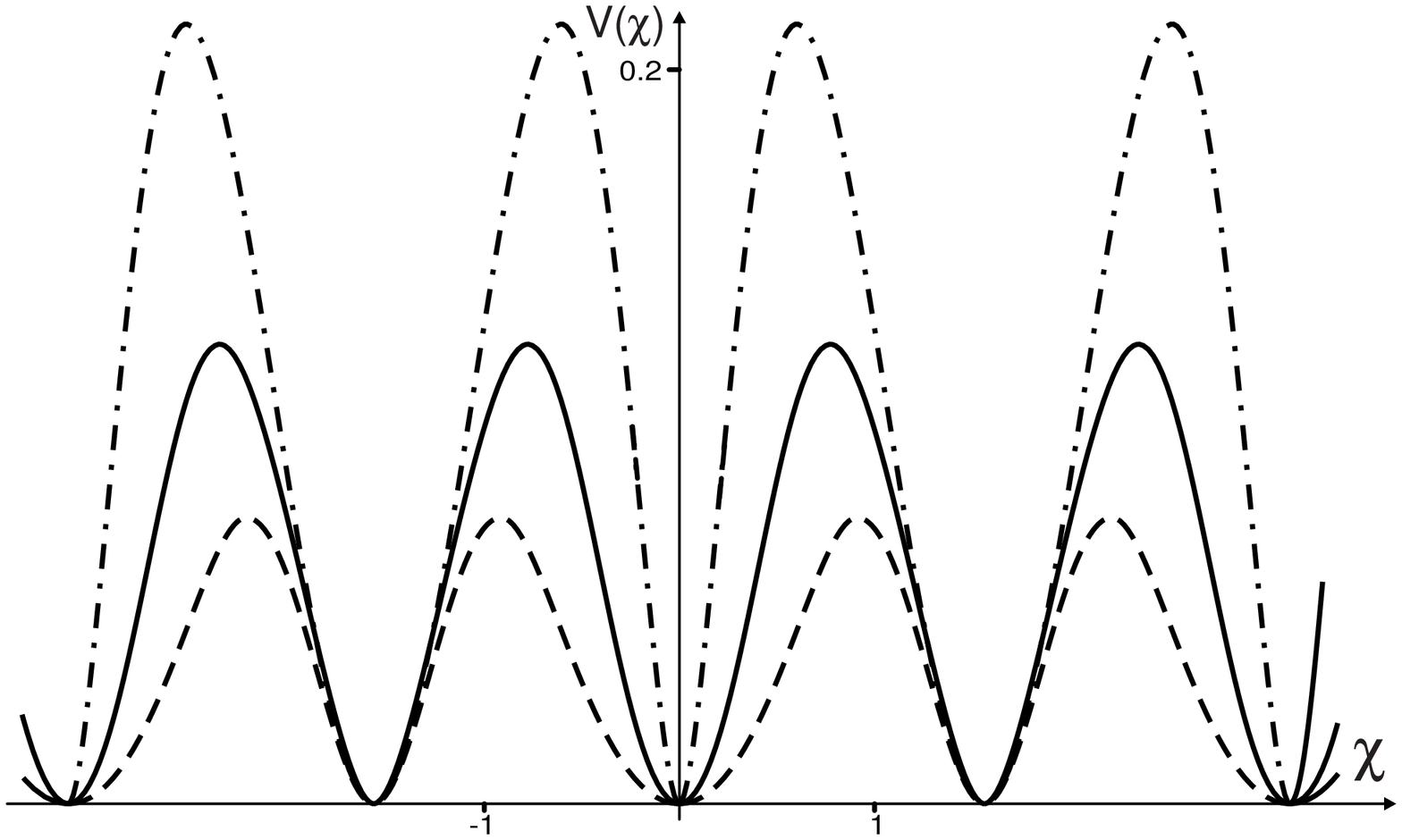}
\hspace{.5cm}
\includegraphics[{width=7cm}]{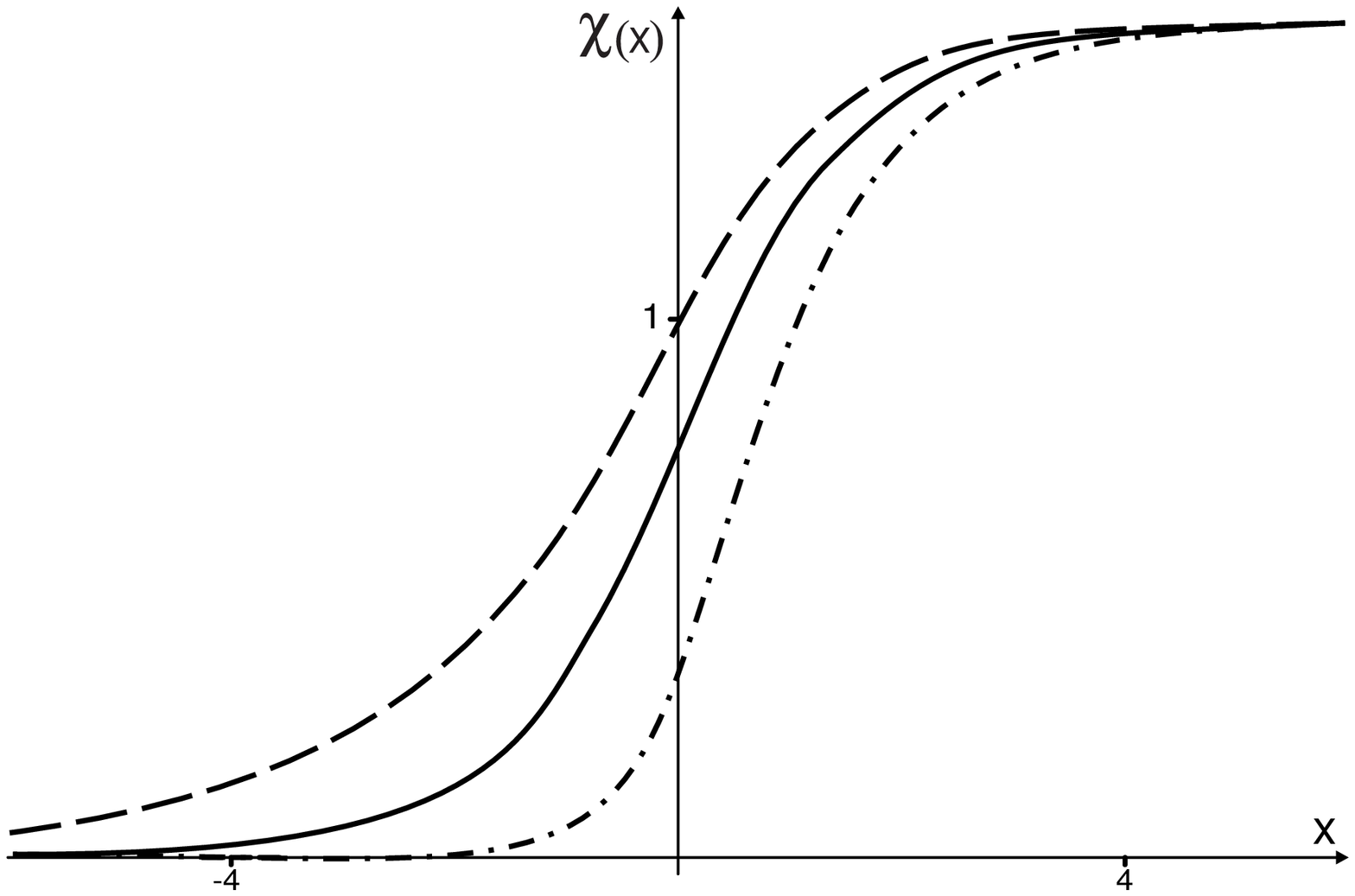}
\vspace{0.3cm} \caption{The potential (\ref{potdsg1}) (left) and  the corresponding kink \eqref{kink1} (right) for $k=0$,
for the cases $n=2/3$ (dashed line), $n=2$ (solid line), and $n=4$ (dash-dotted line).}
\end{figure}

The static kink-like solutions of the model (\ref{potdsg1}) are given by
\begin{equation}\label{kink1}
{\chi}_{n}^{\pm}(x)=k\pi\pm\arcsin\left[\left(\frac12(1\mp\tanh(x))\right)^{1/n}\right]  ,
\end{equation}
where $k=0,1,2,3,...$, 
which correspond to the deformation of the kinks of the $\phi^4$ model. These potentials and corresponding solutions are plotted
in Fig.~2, for the values $n=2/3,2$, and $4$. Also, the energy densities are given by
\be\label{ened1}
\rho^{\pm}(x)=\frac1{n^2}\frac{\left(1\mp\tanh(x)\right)^{2/n-2}\left(1-\tanh^2(x)\right)^2}{2^{2/n}-\left(1-\tanh(x)\right)^{2/n}} .
\ee
In fig.~3 we plot the energy densities for the values $n=2/3,2$, and $4$, and the energy as a function of $n;$ and there we see that the energy goes to $\pi^2/12$  for $n\to0,$ and vanishes in the limit $n\to\infty.$

\begin{figure}[ht!]
\includegraphics[width=5cm]{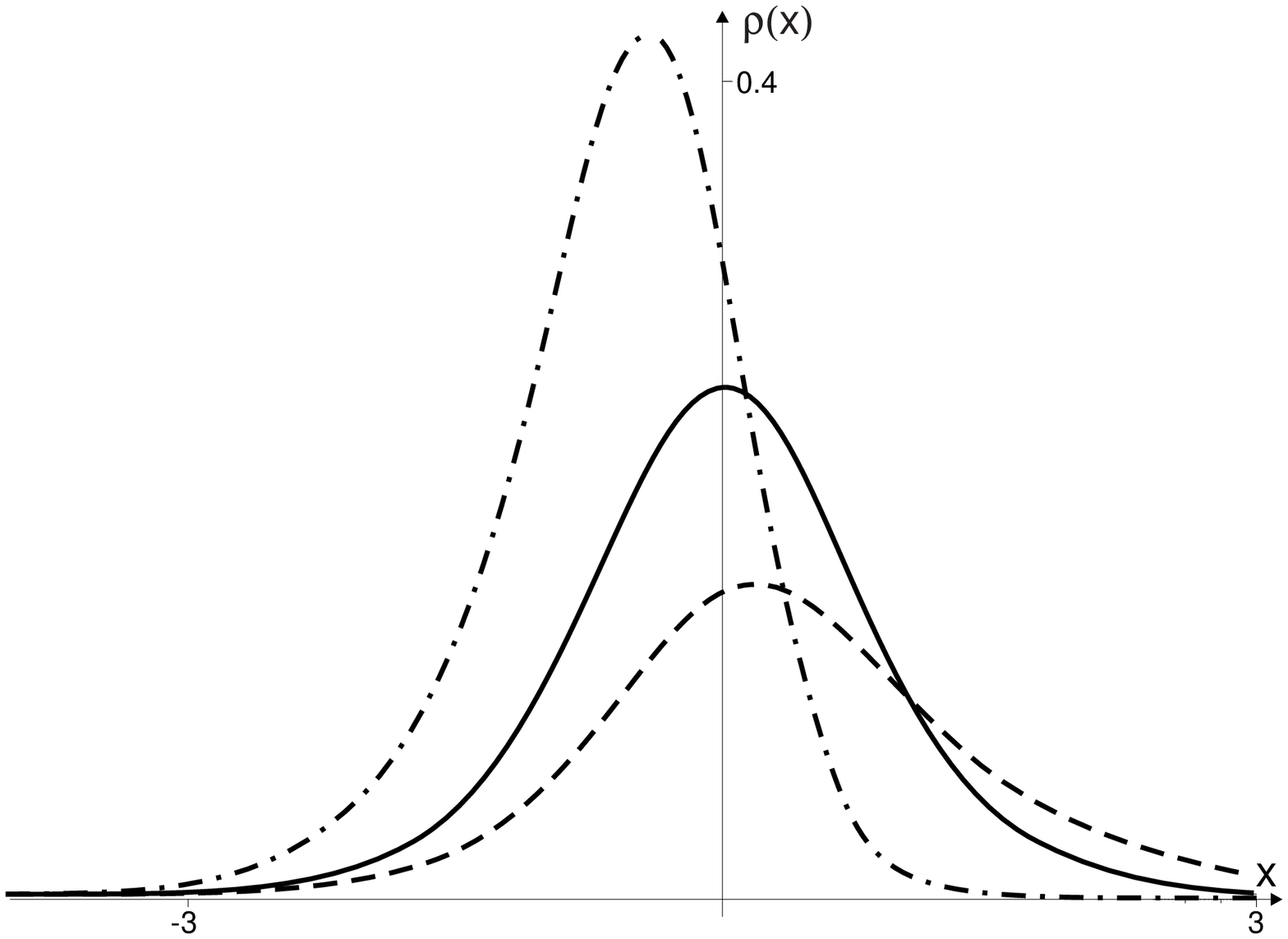}
\includegraphics[width=5cm]{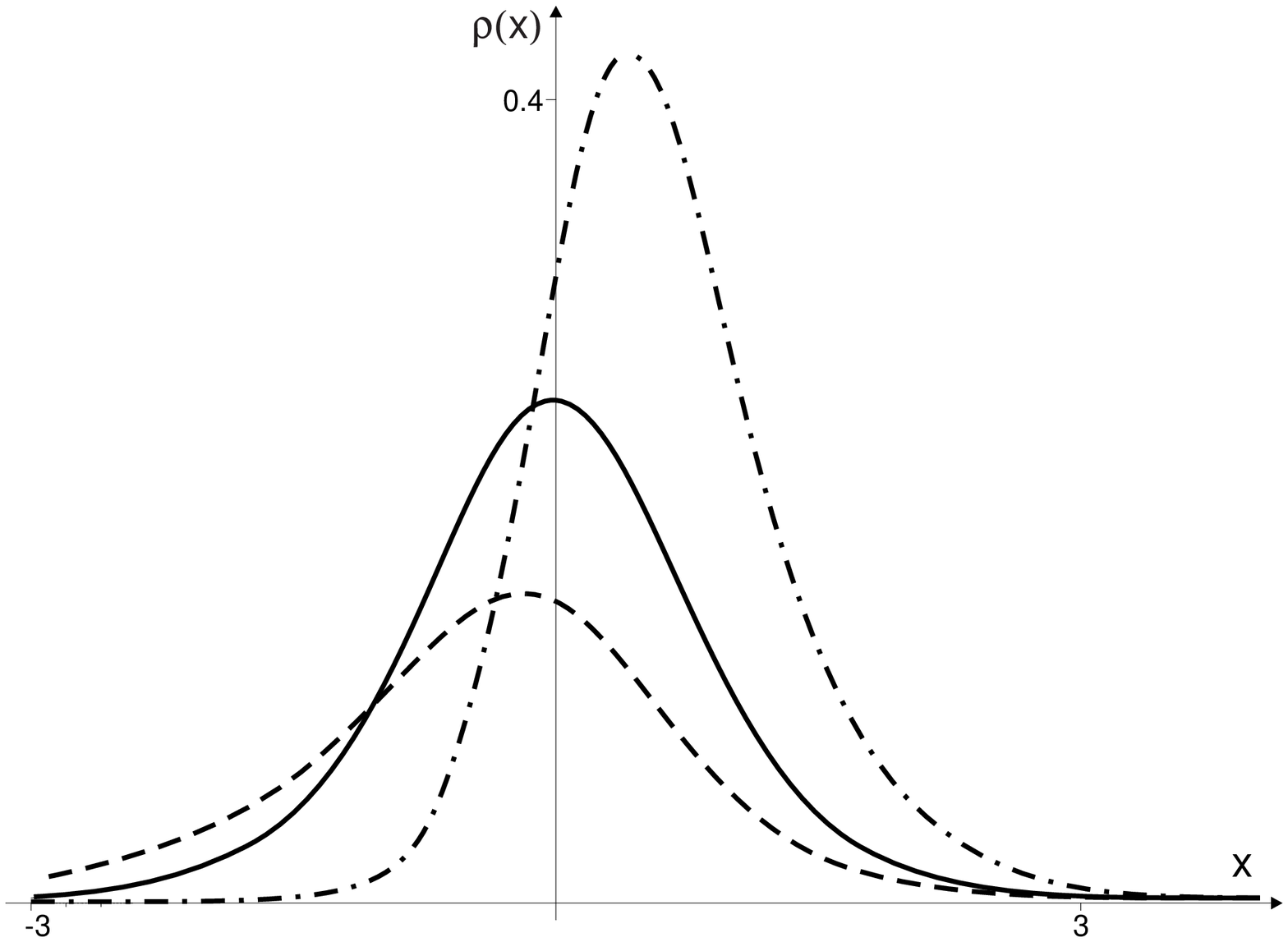}
\includegraphics[width=5cm]{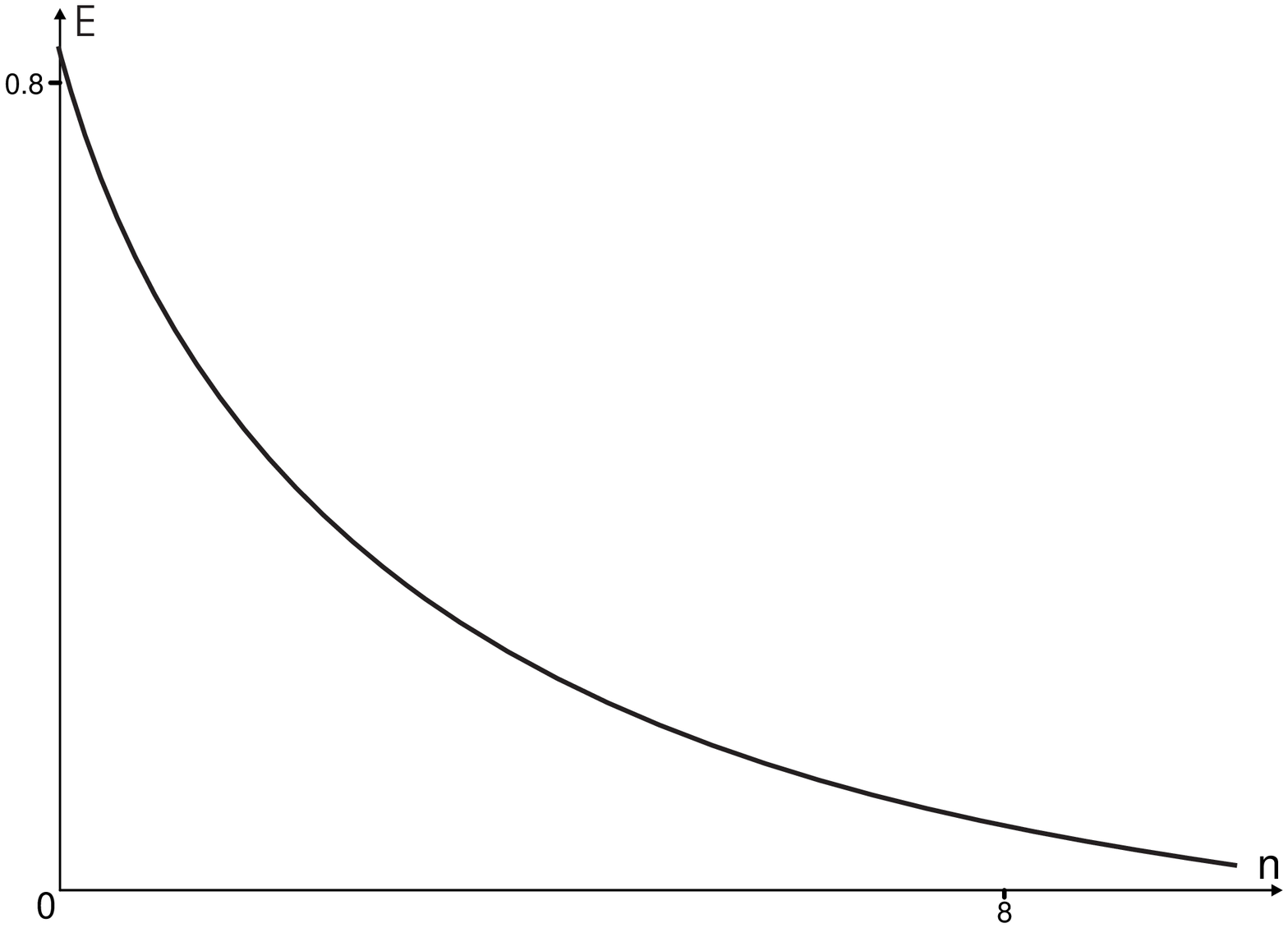}
\vspace{0.3cm} \caption{The energy density (\ref{ened1}) for $\rho^{+}$ (left) and $\rho^{-}$ (center),
for the values $n=2/3$ (dashed line), $n=2$ (solid line), and $n=4$ (dash-dotted line), and the energy (right) as a function of $n$.}
\end{figure}

For both solutions, the potential of the Schr\"odinger-like equation can be cast to the form
\ben\label{qm1}
U(x)&=&-4\;a(x)\left(1-2a(x)\right)-\frac{12\;n\;a(x)\left(1-a(x)\right)-8\;a(x)\left(2-a(x)\right)}{n^2\left(1-a(x)^{2/n}\right)}+\frac{12\left(1-a(x)\right)^2}{n^2\left(1-a(x)^{2/n}\right)^2}
\een
where $a(x)=1/2\left(1\pm\tanh(x)\right)$, with $+$ ($-$) corresponding to the $\chi^{-}(x)$ ($\chi^{+}(x)$) solution. These potentials are plotted in Fig.~4, for the values $n=2/3,2$, and $4$. 

\begin{figure}[ht]
\includegraphics[{width=8cm}]{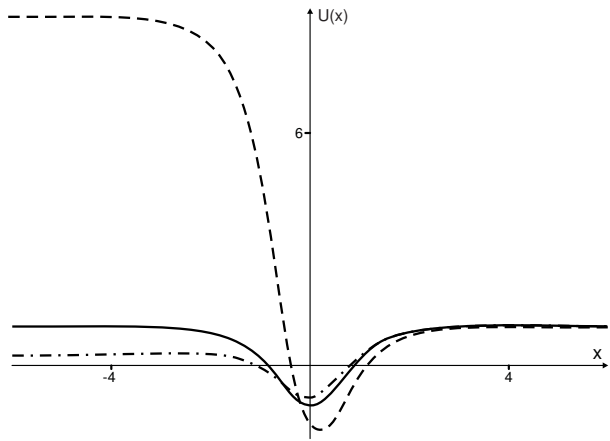}
\vspace{0.3cm} \caption{Plot of the potential of the Schr\"odinger-like equation \eqref{qm1},
for the cases $n=2/3$ (dashed line), $n=2$ (solid line), and $n=4$ (dash-dotted line).}
\end{figure}

The asymmetry in $U(x)$ in Fig.~4 is due to the different masses of the excitations around the distinct minima. For $\chi^{\pm}_n(x)$, we find
\be
U(\pm\infty) = 1 \,\,\,\,\,\,\,\,\, {\rm{and}}
\,\,\,\,\,\,\,\,\,U(\mp\infty) = \frac{4}{n^2} .
\ee
This asymmetry is also related to the asymmetry present in the energy densities $\rho^+$ and $\rho^-$ which are plotted in Fig.~3. It is also known to appear in the $\phi^6$ model described by $W_{\phi}=\phi(1-\phi^2)$ -- see e.g., Ref.{\cite{lohe}}.

If we consider $n=2+\delta$, for $|\delta|<<1$, we can approximate the potential (\ref{potdsg1}) by
\be
{\wt V}(\chi)=\frac18\sin(2\chi)^2-\frac12\left(\sin(\chi)^2\ln\left(\sin(\chi)^2\right)-\frac14\sin(2\chi)^2\right)\;\delta\;,
\ee
and the corresponding kink-like solutions (\ref{kink1}) by
\be
{\wt \chi}^{\pm}_n(x)=\chi^{\pm}_2(x)\mp\frac14e^{\mp x}\ln\left(\frac12\left(1\mp\tanh(x)\right)\right)\;\delta\;.
\ee
The energy of these solutions now becomes
\be
{\wt E}_n= E_2-\left(\frac12-\frac{\pi^2}{24}\right)\,\delta ,
\ee
for $\delta<<1.$ In this case the model is very close to the sine-Gordon model, and this may be useful for both semiclassical and phenomenological investigations. 

The class of models considered above can be seen as deformation of the sine-Gordon model, widening ($n\leq2$) or marrowing ($n\geq2$) the width around alternate minima of the sine-Gordon model ($n=2$) -- see Fig.~2. This changes the classical mass associated to these minima, and would modify the way the quantum corrections appear. As we already know from the $\phi^6$ model \cite{lohe}, however, the asymmetry of the potential of the Schr\"odinger-like equation turns it hard to investigate the quantum effects.

\subsection{Type-II models}

We consider another class of models, obtained with the deformation function
\be\label{c3Sine3phi_4}
f_p(\chi)=\cos\left(\chi^p\right) .
\ee
We use it to deform the $\phi^4$ model, to get to the new deformed potentials
\begin{equation}\label{potdsg6}
{\wt V}(\chi)=\frac1{2p^2}\chi^{2-2p}\,\sin^2(\chi^p) ,
\end{equation}
where $p$ can be integer positive, or a rational odd/odd, or even/odd number.
For a given value of $p$, the potential (\ref{potdsg6}) has infinite minima located at
\be
{\bar\chi}^p_0 = \pm |k\pi|^{\frac1p}
\ee
where $k$ is a natural number. The nonvanishing maxima satisfy the relation
\be
{\bar\chi}^p_m = \frac{p-1}{p} \tan\left({\bar\chi}^p_m\right) .
\ee
In this case, the maximum can be approximated by the midpoint between two minima, and this can be further approxiomated to ${\tilde\chi}^p_m\approx |(k+1/2)\pi|^{1/p},$ such that 
\be
V({\tilde \chi}^p_m)\approx \frac{\pi^{2(1-p)/p}}{2p^2}\left(k+\frac12\right)^{2(1-p)/p}
\ee
and the ratio between two consecutive maxima has the form $(1+1/(2k+1))^{2(1-p)/p}.$
 
We see that for $p=1$, the potential (\ref{potdsg6}) reduces to the sine-Gordon potential. For $p\neq 1$, the potential is an oscillatory non-periodic function of $\chi$ having two distinct features, depending on the value of the parameter $p$. For $p\in(0,1)$, the amplitude of oscillation of ${\wt V}(\chi)$ increases with increasing $|\chi|$ and for $p\in(1,\infty)$, the amplitude decreases for increasing $|\chi|$, as we illustrate in Fig.~5. With exception of
the central minimum, all other minima have the same concavity, given by $d^2{\wt V}/d\chi^2=1$. For the minimum at $\chi=0$, we find
$d^2{\wt V}/d\chi^2=1/p^2$. Thus, the concavity at the symmetric minimum ${\bar\chi}=0$ is greater or lesser then the concavity at every other
(asymmetric) minima for $p\in(0,1)$ or $p\in(1,\infty),$ respectively.
In Fig.~5 we plot the potential (\ref{potdsg6}) for the cases $p=2/3,2,$ and $4$ and the corresponding kink \eqref{kink2} for $k=0,$ and $1$.

\begin{figure}[ht]
\includegraphics[{width=6cm}]{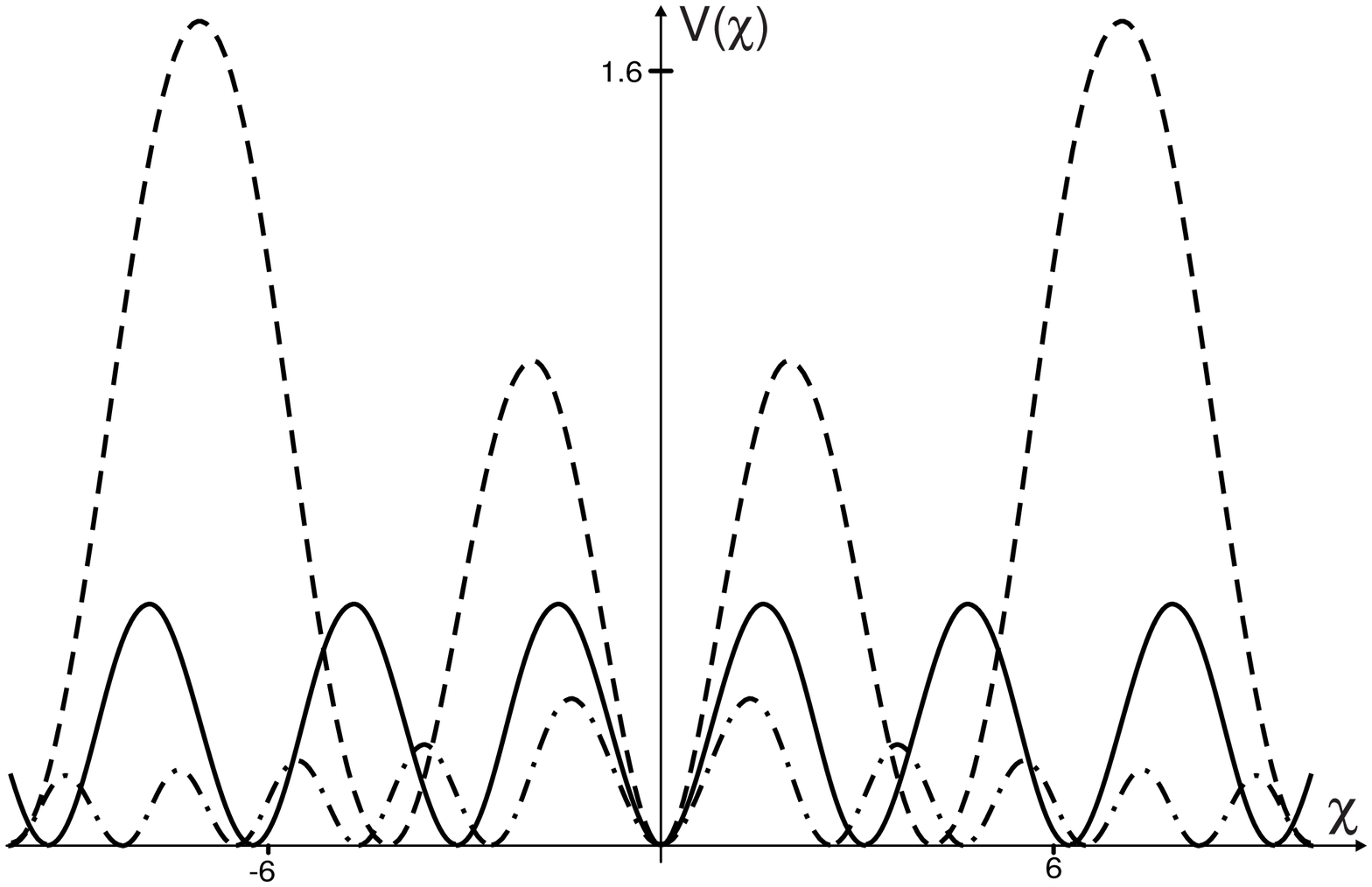}
\includegraphics[{width=5cm}]{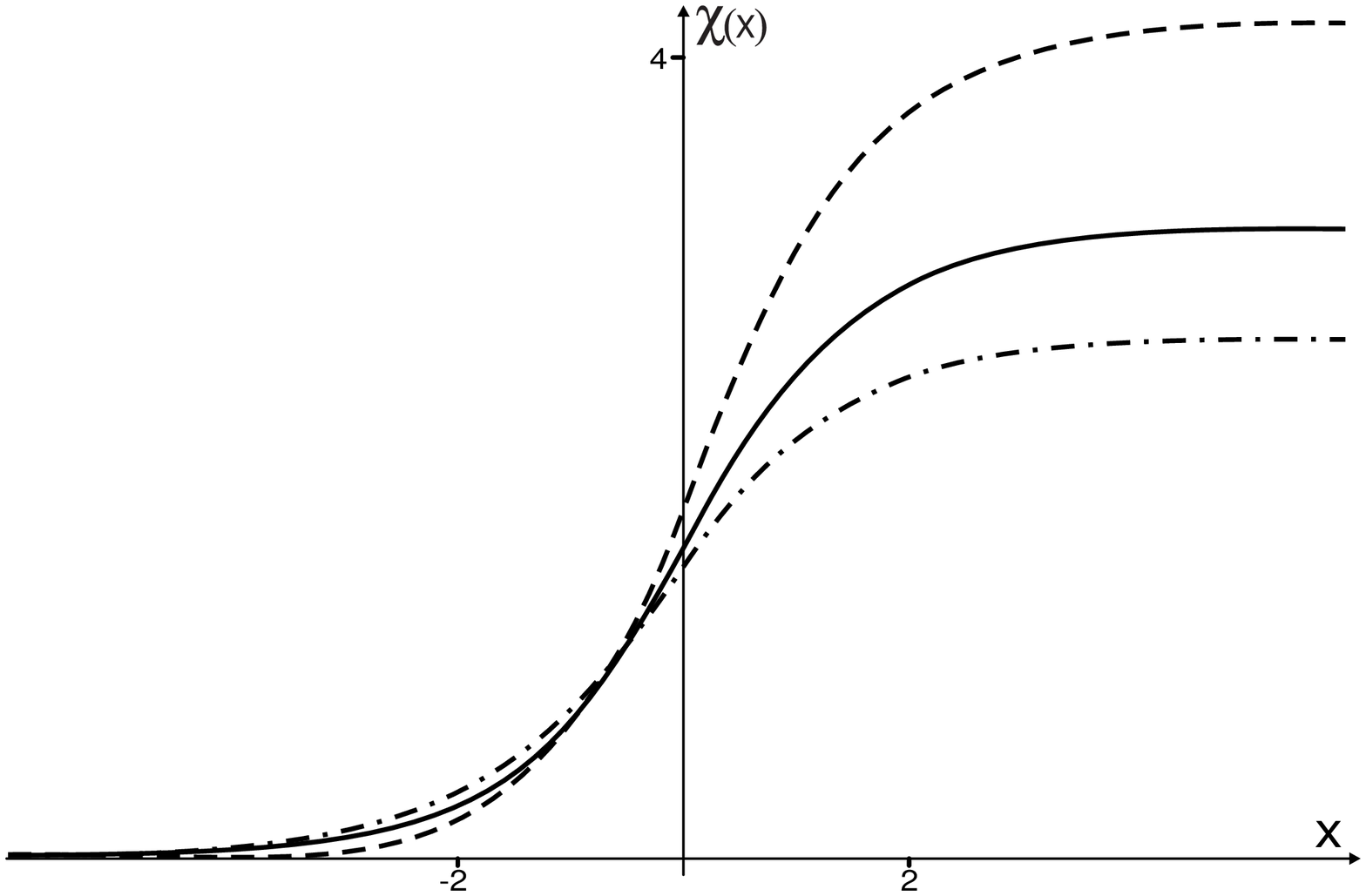}
\includegraphics[{width=5cm}]{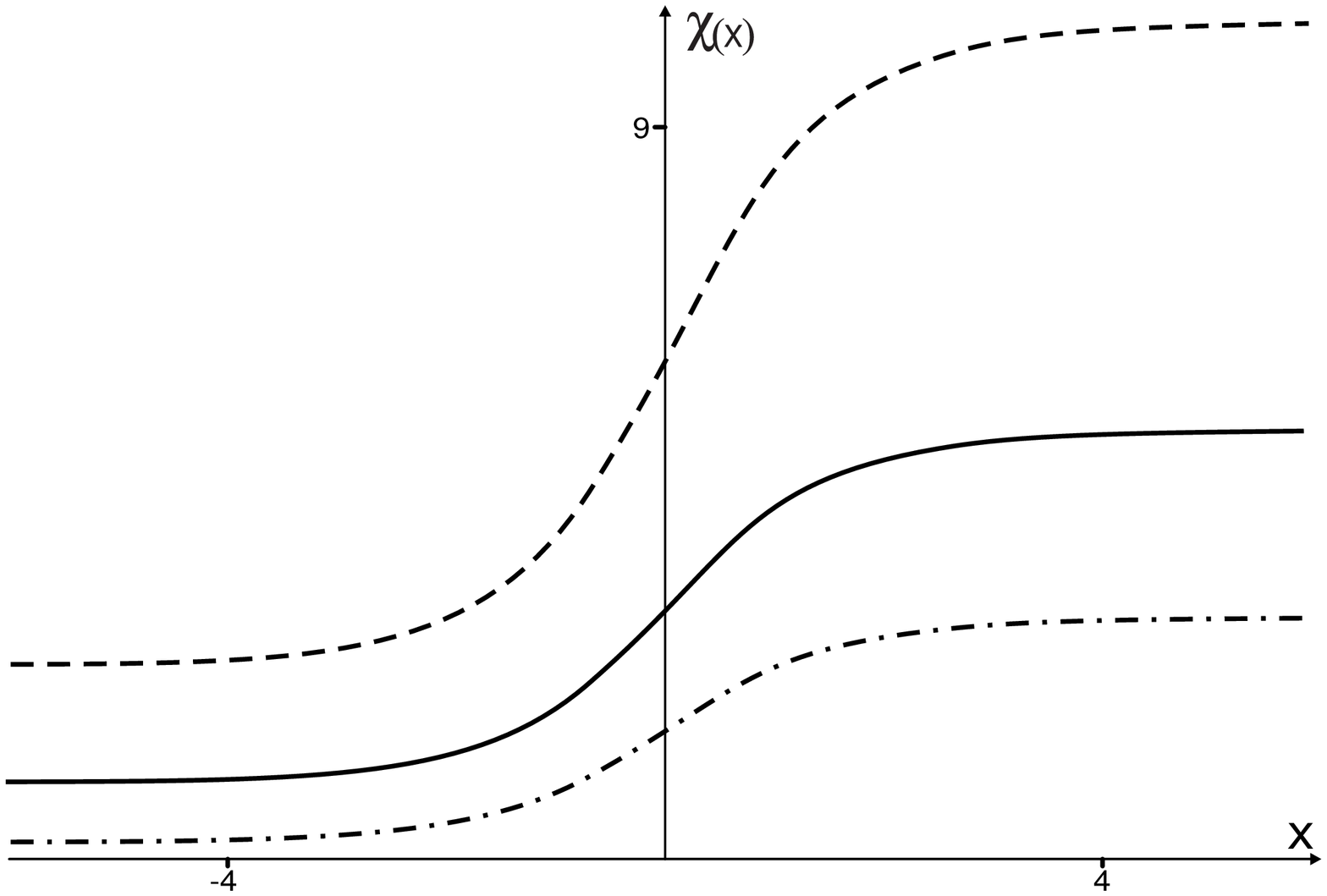}
\caption{The potential (\ref{potdsg6}) (left), the corresponding kink \eqref{kink2} for $k=0$ (center), and for $k=1$ (right), for the cases  $p=4/5$ (dashed line), $p=1$ (solid line), and $p=6/5$ (dash-dotted line).}
\end{figure}

For an arbitrary value of $p$, the superpotentials associated with the potentials (\ref{potdsg6}) do not have a simple form, but can be
expressed in terms of hypergeometric functions as
\be
W_p(\chi)=\frac{\chi^2}{2p} \mbox{ }_1
F\left(\frac1p;\frac32,\frac{p+1}p; -\frac{\chi^{2p}}{4} \right),
\ee
and the energy of the localized solutions connecting adjacent minima labeled by $k$ and $k+1$ has the form
\be
E_p=|W_p(\bar\chi^p_0(k))-W_p(\bar\chi^p_0(k+1))|.
\ee
For some choices of the value of $p$, however, simpler superpotentials can be found. For example, with $p=2$ we find
\be
W_2(\chi)=\frac14 {\rm Si}(\chi^2) ,
\ee
where ${\rm Si}(z)$ is the sine integral function. In this case the energy is
\be
E_2=\frac{(-1)^k}{4}\left(\;{\rm Si}((k+1)\pi)\;-\;{\rm Si}(k\pi)\;\right) .
\ee
For $p=2/3$, the superpotential becomes
\be
W_{2/3}(\chi)=\frac94 \left(\sin(\chi^{2/3}) -\chi^{2/3}\,\cos(\chi^{2/3}) \right),
\ee
and we find the energy
\be
E_{2/3}=\frac{9}{4}\left(2k+1\right)\pi ,
\ee
which depends linearly on $k$. For $p=1/3,$ the energy is
\ben
E_{1/3}=9\left(2\pi^4{k}^{4}+4\pi^4{k}^{3}+ 6\pi^2\left({\pi}^{2}-4 \right) {k}^{2}+
4\pi^2\left( {\pi}^{2}-6\right)k+\pi^4-12\pi^2+48\right)
\een
and it now depends on $k$ as a fourth-order polynomial. For the cases $p=1/5$ and $p=2/5$, the energy $E_{p}$ are polynomials in $k$ of order $8$ and $3$, respectively.

The topological solutions can be found by considering the first order equation
\be
\frac{d\chi}{dx}=\frac1{p}\frac{\sin(\chi^p)}{\chi^{p-1}} .
\ee
Using the inverse of the deformation function (\ref{c3Sine3phi_4}) and the kink solution of the $\phi^4$ model, we obtain the solutions
\bes\label{kink2}
\ben
\chi_k(x)&=&\pm\left((k+1)\pi-\arccos(\tanh(x))\right)^{1/p} , \\
\chi_{\bar k}(x)&=&\pm\left({k}\pi+\arccos(\tanh(x))\right)^{1/p} ,
\een
\ees
where $k=0,1,2,3,\dots\,$; the integers $k$ and ${\bar k}$ label the set of kink and anti-kink solutions, which appear in the sine-Gordon-like model described by the potential (\ref{potdsg6}).

The respective energy densities are given by
\bes
\ben\label{ened2}
\rho_k(x)&=&\frac1{p^2} \frac{\sech^2(x)}{\left((k+1)\pi -
\arccos(\tanh(x))\right)^{2-\frac2p}} , \\
\rho_{\bar k}(x)&=&\frac1{p^2} \frac{\sech^2(x)} {\left({k}\pi+\arccos(\tanh(x))\right)^{2-\frac2p}} ,
\een\ees
which we plot in Fig.~6 for $k=0,1,$ for some values of $p$ and the energy as function of $p$. The potentials of the Schr\"odinger-like equations associated with these solutions are given by 
\bes\label{qm2}\ben
U_k(x)&=&1-2\sech^2(x) + \left(\frac1p -1\right)
\Big[\left(\frac1p-2\right)
\frac{\sech^2(x)}{a_k^2(x)} -3\frac{\tanh(x)\sech(x)}{a_k(x)} \Big], \\
U_{\bar k}(x)&=&1-2\sech^2(x) + \left(\frac1p -1\right)
\Big[\left(\frac1p-2\right) \frac{\sech^2(x)}{a_{\bar
k}^2(x)}+3\frac{\tanh(x)\sech(x)}{a_{\bar k}(x)}   \Big] ,
\een \ees
where $a_k(x)=(k+1)\pi-{\rm arccos}\left(\tanh(x)\right)$ and $a_{\bar k}(x)=k\pi+{\rm arccos}\left(\tanh(x)\right)$. The potential of the Schr\"odinger-like equation \eqref{qm2} is plotted in Fig.~7, and we notice that they lead to \eqref{sgq} in the limit $p\to1,$ as expected. 

The asymmetry in $U(x)$ is due to the different masses of the excitations around the distinct minima. For the solutions $\chi_{k,\bar k}(x)$, we find for $k=0$
\be
U_{k,\bar k}(-\infty) = \frac1{p^2},1 \,\,\,\,\,\,\,\,\, {\rm{and}}
\,\,\,\,\,\,\,\,\,U_{k,\bar k}(+\infty) = 1 ,\frac1{p^2}\; ,
\ee
and for $k\geq 1$
\be
U_{k,\bar k}(\mp\infty) = 1 .
\ee

\begin{figure}[ht]
\includegraphics[{width=5.5cm}]{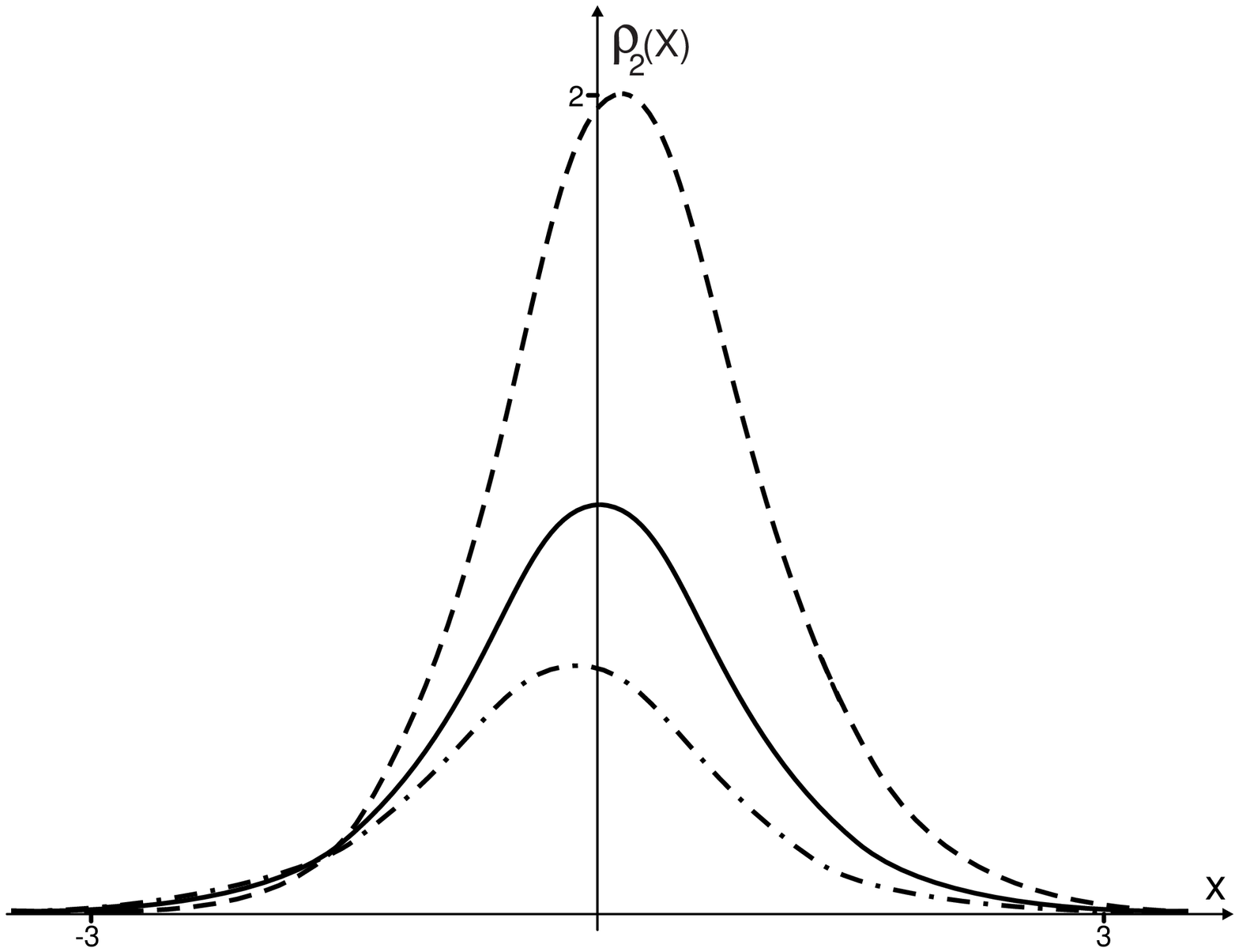}
\includegraphics[{width=5.5cm}]{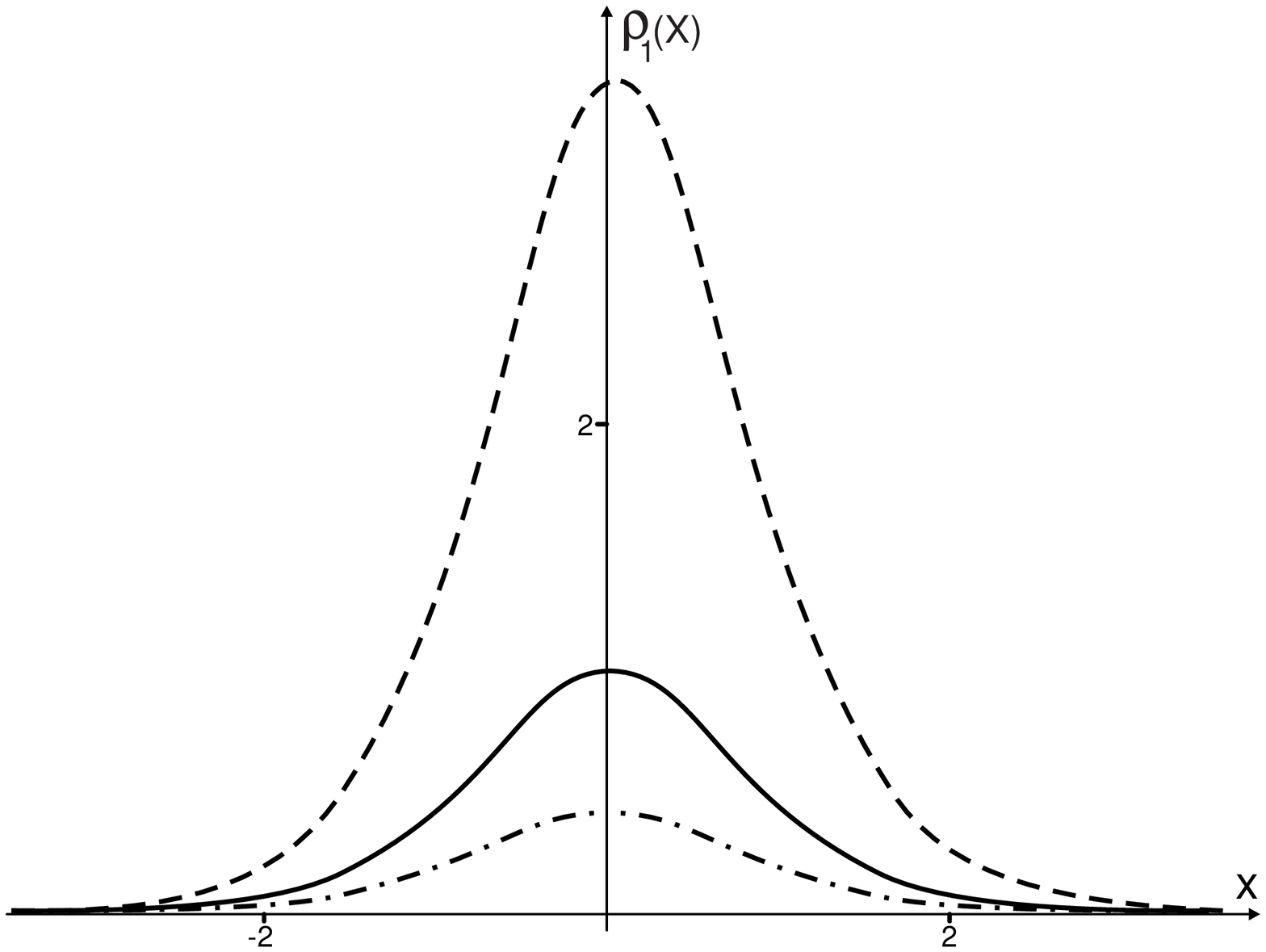}
\includegraphics[{width=6cm}]{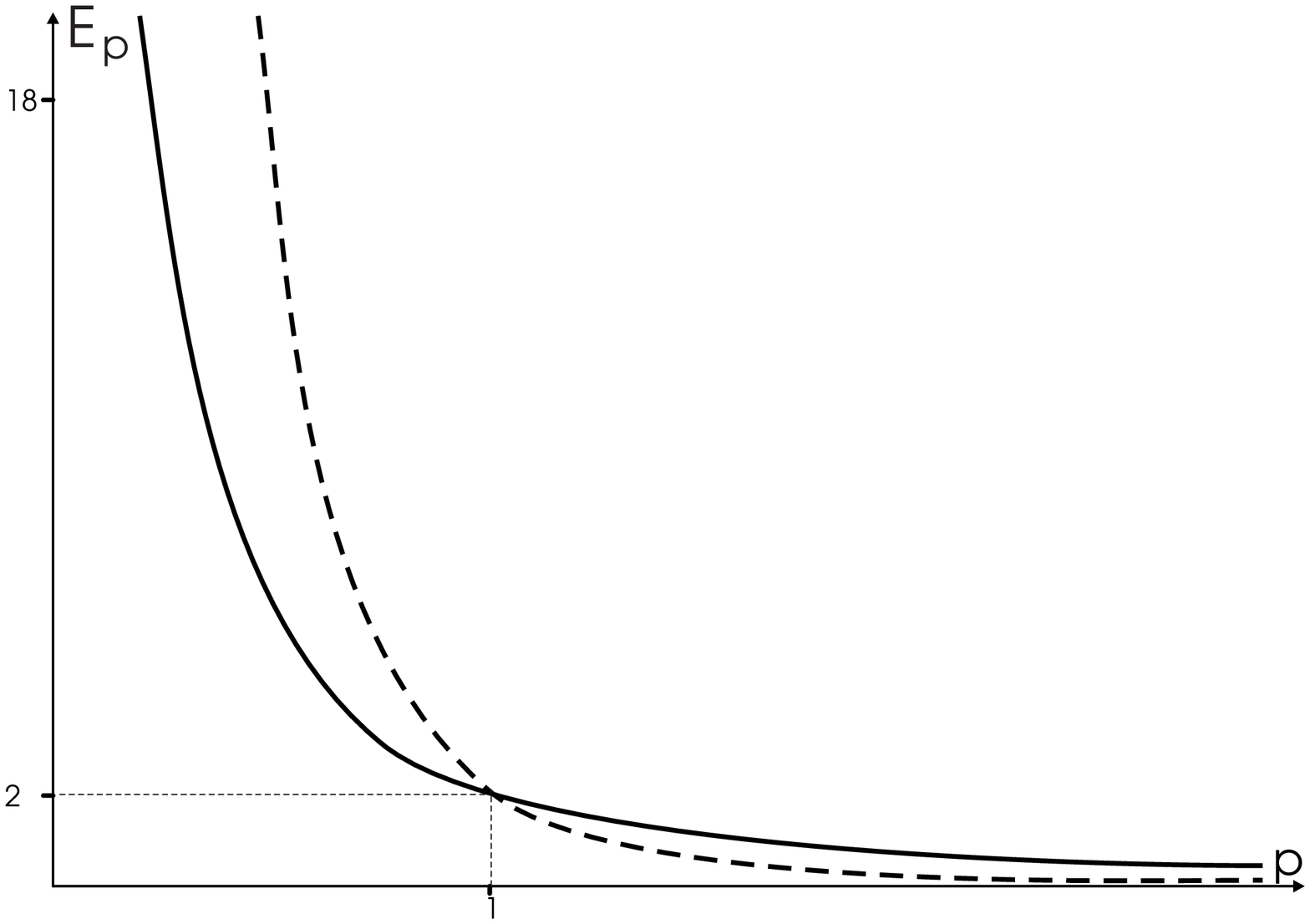}
\caption{The energy density (\ref{ened2}) for $k=0$ (left), and for $k=1$ (center), for the cases  $p=4/5$ (dashed line), $p=1$ (solid line), and $p=6/5$ (dash-dotted line), and the energy (right) as a function of $p$ for $k=0$ (solid line) and for $k=1$ (dashed line).}
\end{figure}

\begin{figure}[ht]
\includegraphics[{width=6cm}]{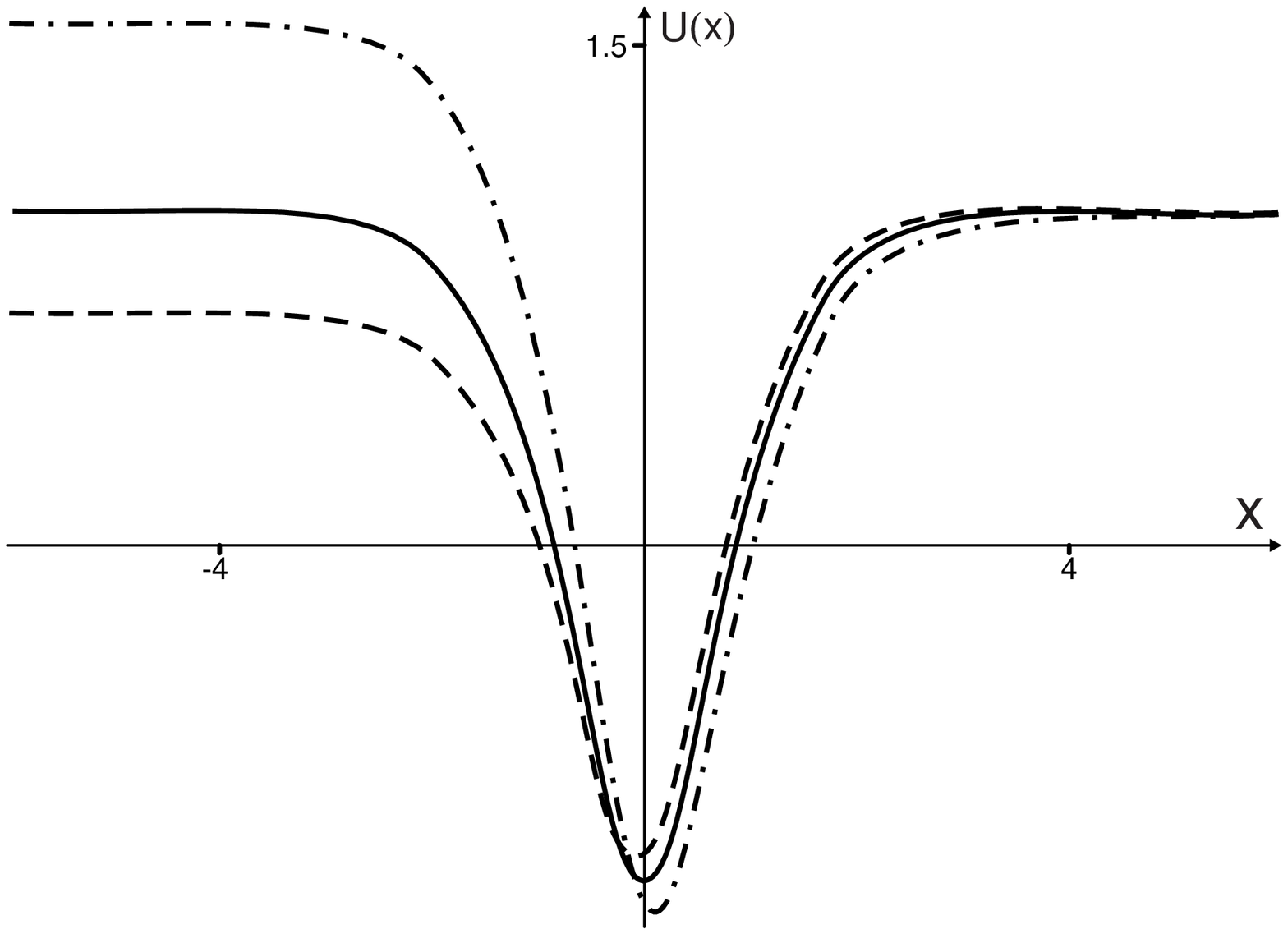}
\hspace{.5cm}
\includegraphics[{width=6cm}]{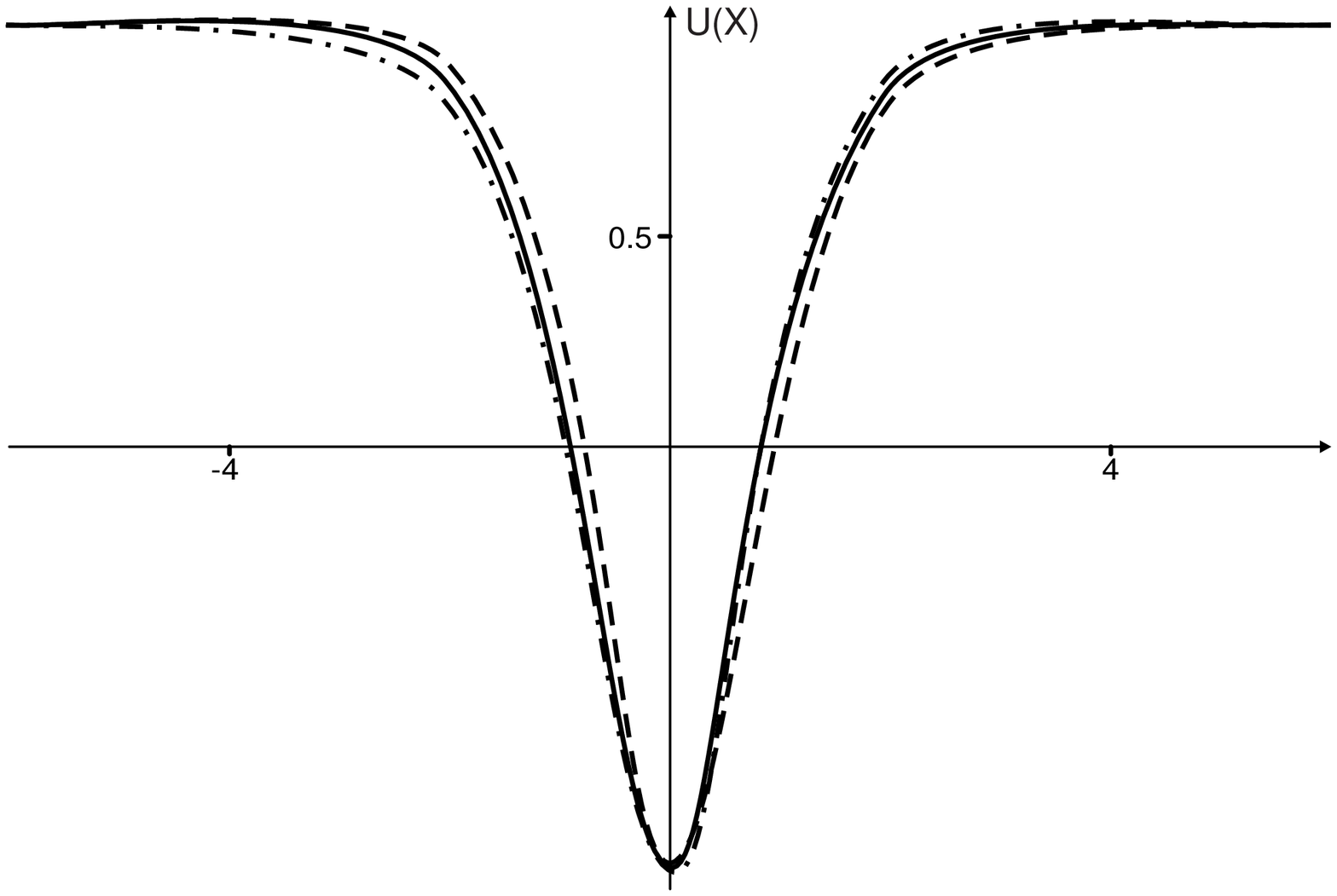}
\caption{Plot of the potential of the Schr\"odinger-like equation \eqref{qm2} for $k=0$ (left), and for $k=1$ (right), for the cases  $p=4/5$ (dashed line),  $p=1$ (solid line), and $p=6/5$ (dash-dotted line).}
\end{figure}

In this new class of models, the asymmetry problem which has appeared in the form case is still present, although now it is more explicit at the central minimum, involving the two first topological sectors. All the other sectors are now {\it almost} symmetric, much easier to be investigated -- indeed, as Fig.~7 shows,
it is very well approximated by the potential which appears in the sine-Gordon case.

The model \eqref{potdsg6} can also have $p$ negative. We plot in Fig.~8 the potential for $p=-1,$ and there we see that the amplitude of the potential oscillates,
decreasing as the field decreases. In Fig.~9 we also plot the $k=1$ profile, energy densities, Schroedinger-like potentials and the total energy of the defect solutions. We note that the energy starts at zero, reaches its maximum at $p=-2,$ and returns to zero as $p$ tends to large negative values. 

\begin{figure}[ht]
\includegraphics[{height=3.5cm,width=3.5cm}]{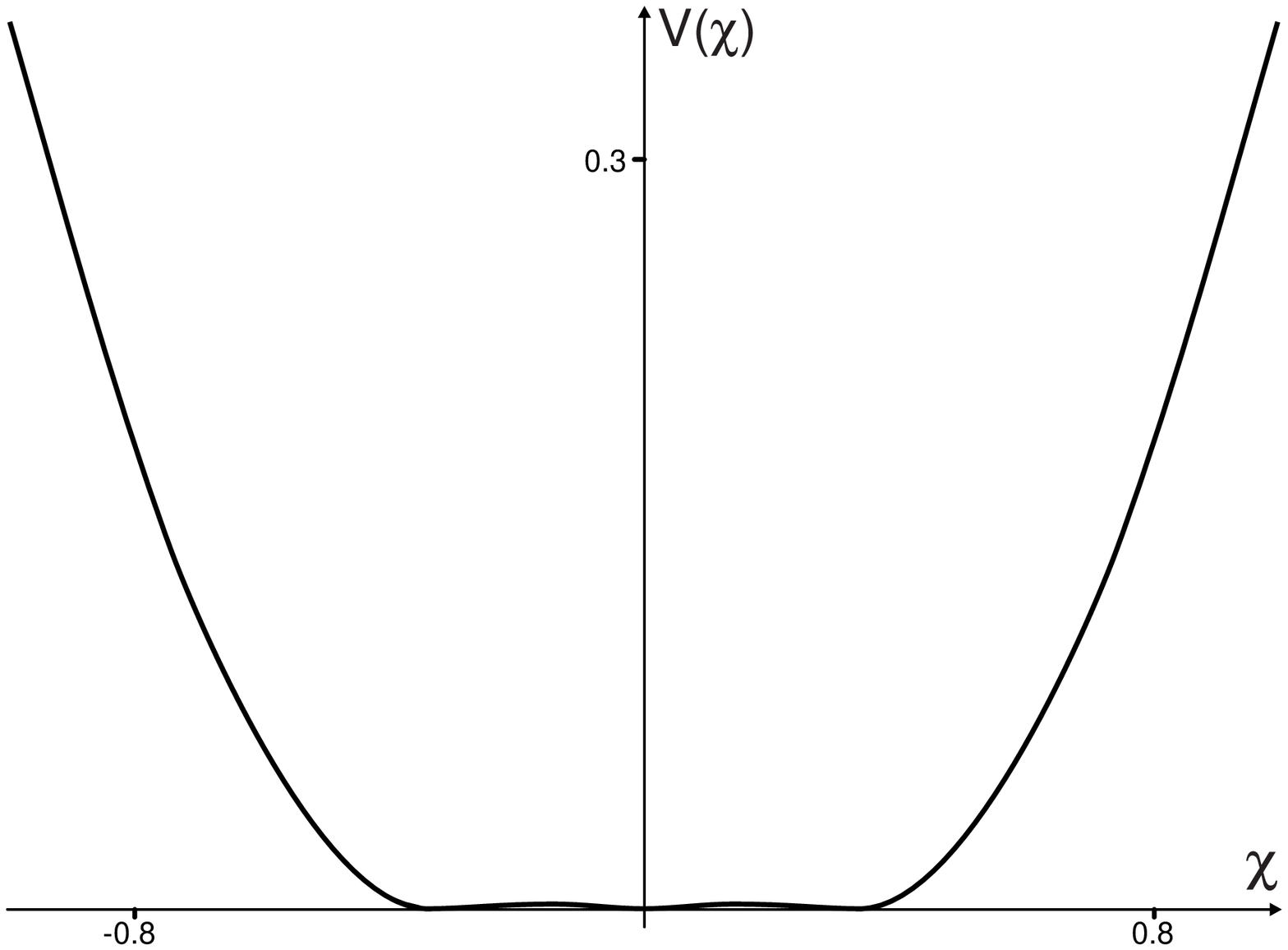}
\hspace{0.5cm}
\includegraphics[{height=3.5cm,width=3.5cm}]{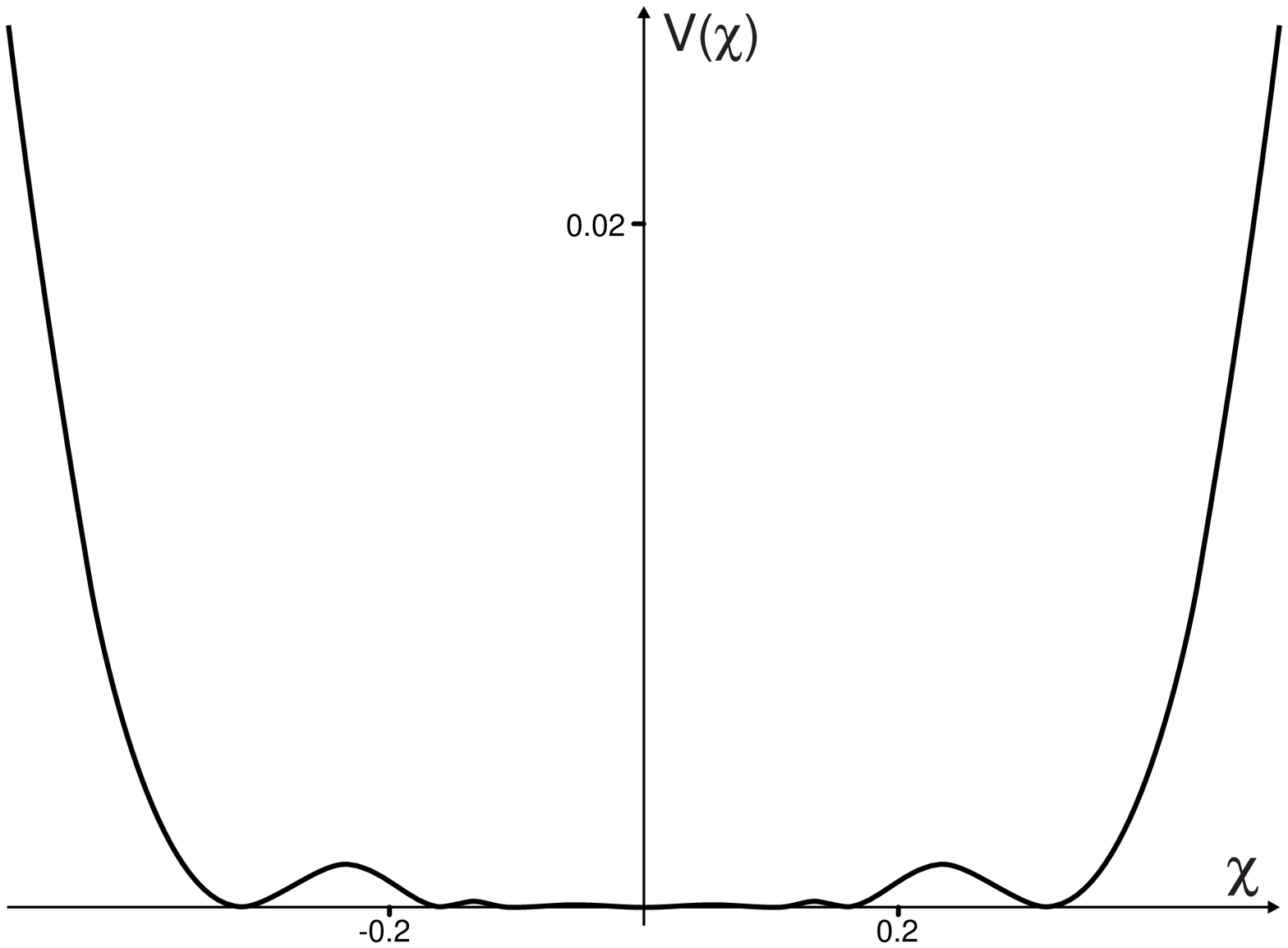}
\hspace{0.5cm}
\includegraphics[{height=3.5cm,width=3.5cm}]{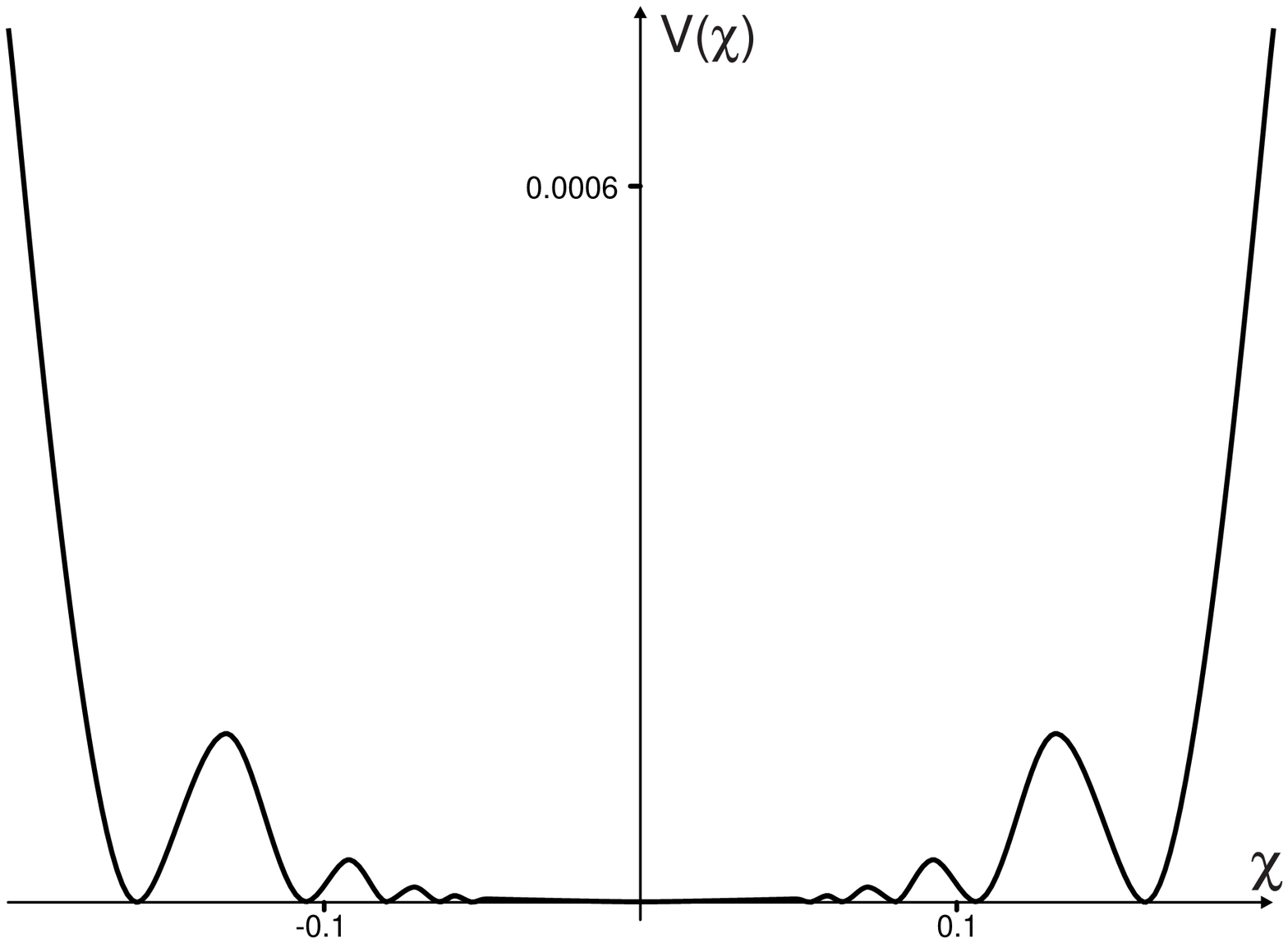}
\hspace{0.5cm}
\includegraphics[{height=3.5cm,width=3.5cm}]{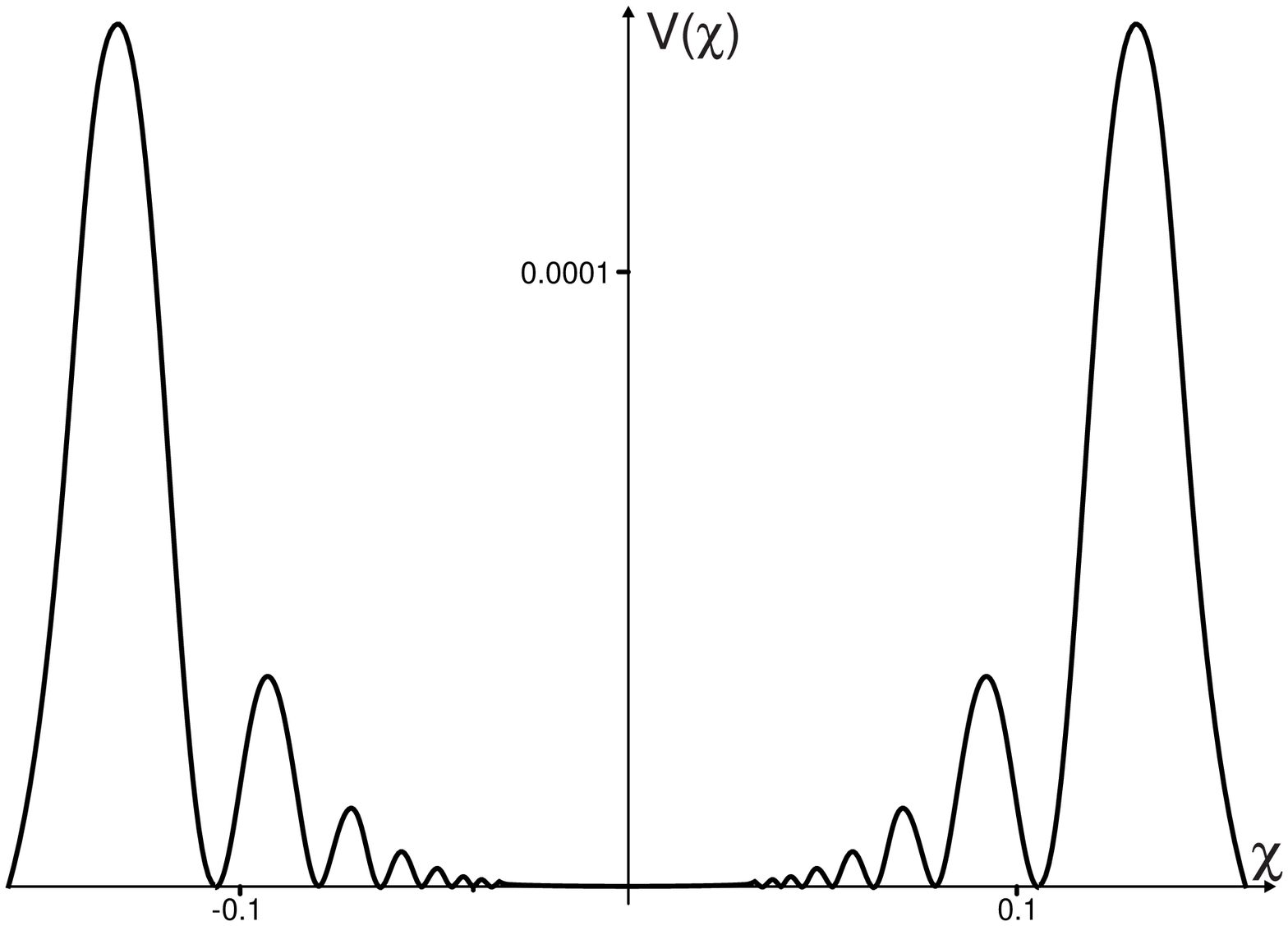}
\caption{The potential (\ref{potdsg6}) for $p=-1,$ depicted for closer and closer distances from the origin.}
\end{figure}

\begin{figure}[ht]
\includegraphics[{height=3.5cm,width=3.5cm}]{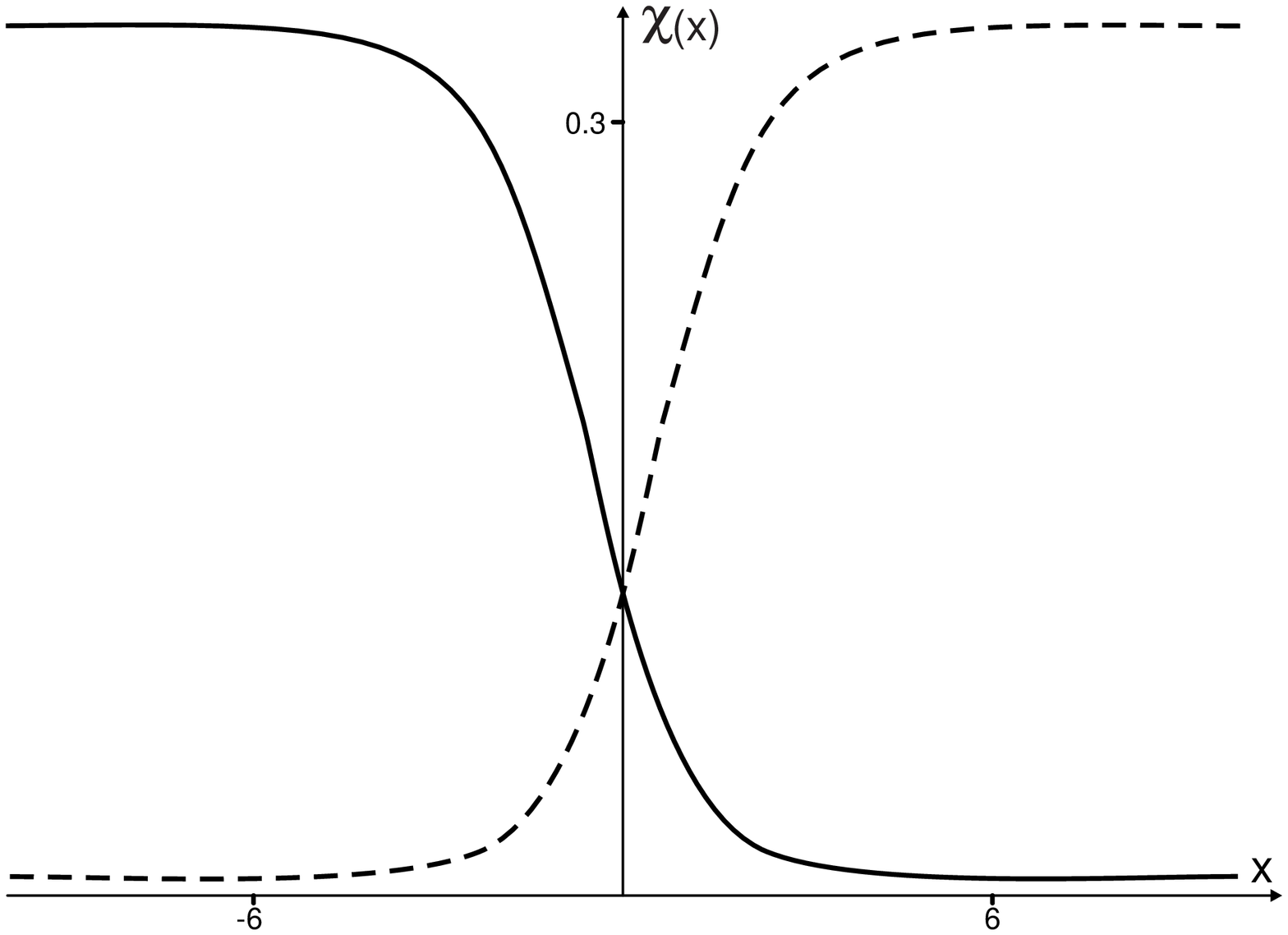}
\hspace{0.5cm}
\includegraphics[{height=3.5cm,width=3.5cm}]{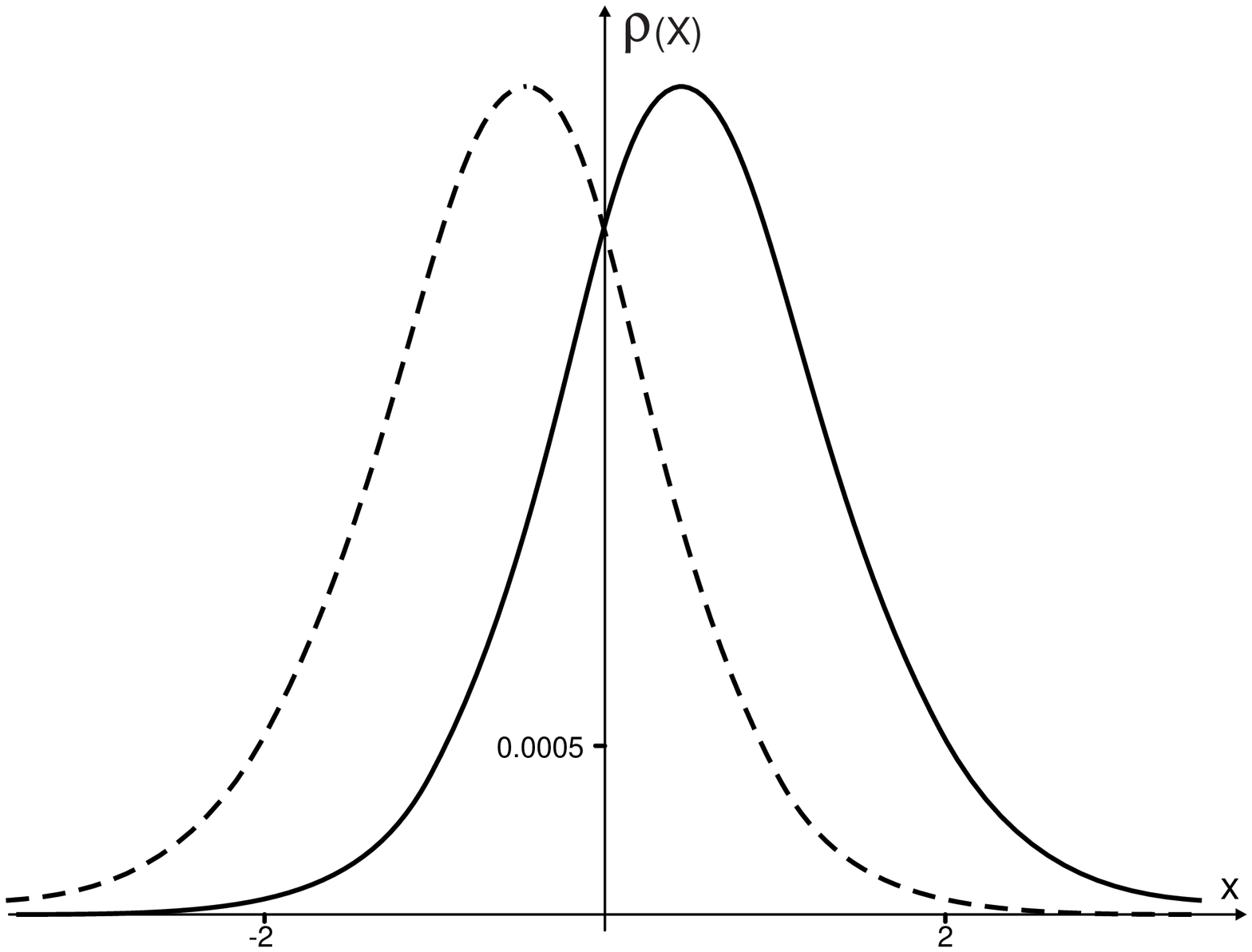}
\includegraphics[{height=3.5cm,width=3.5cm}]{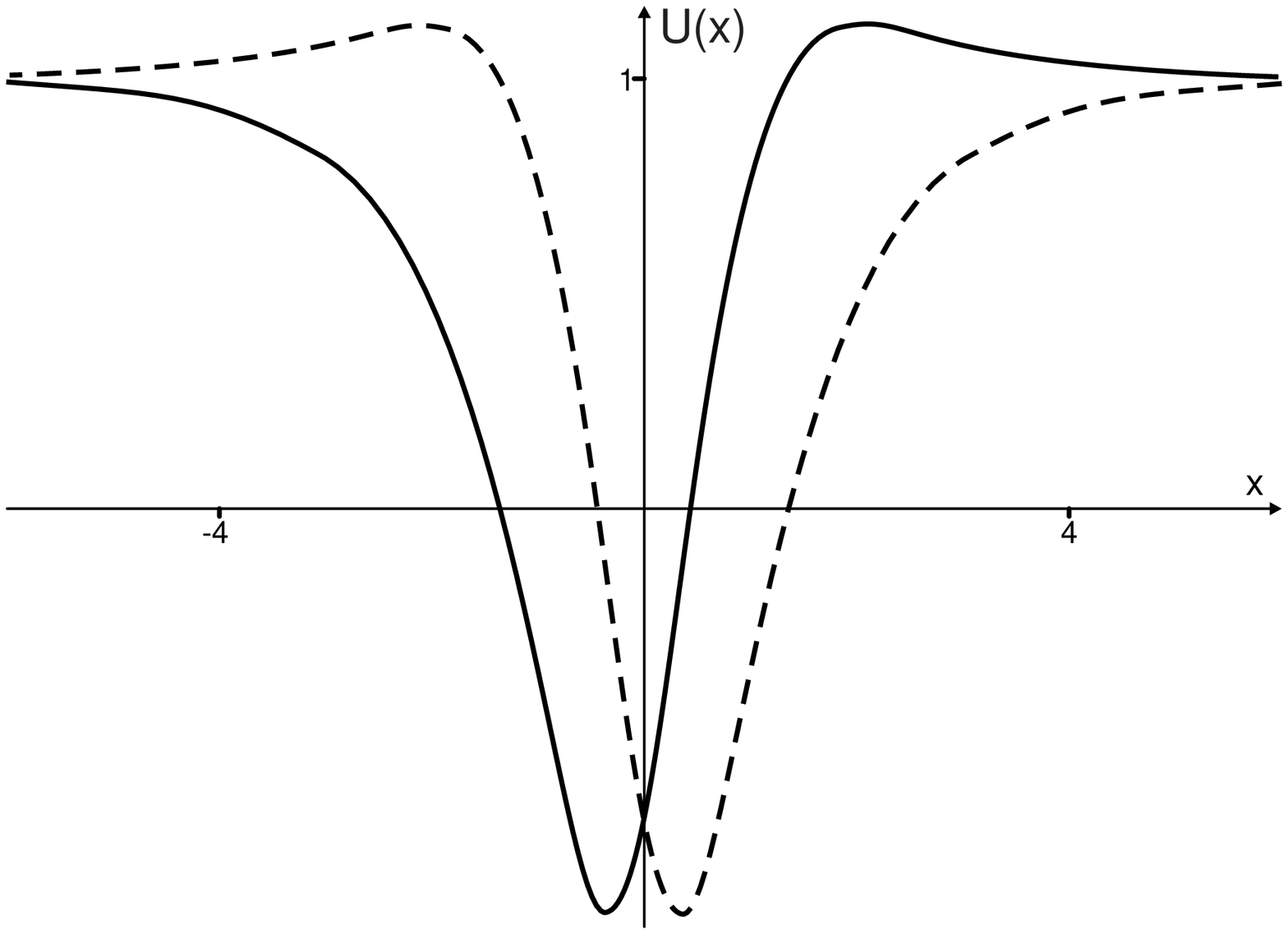}
\includegraphics[{height=3.5cm,width=3.5cm}]{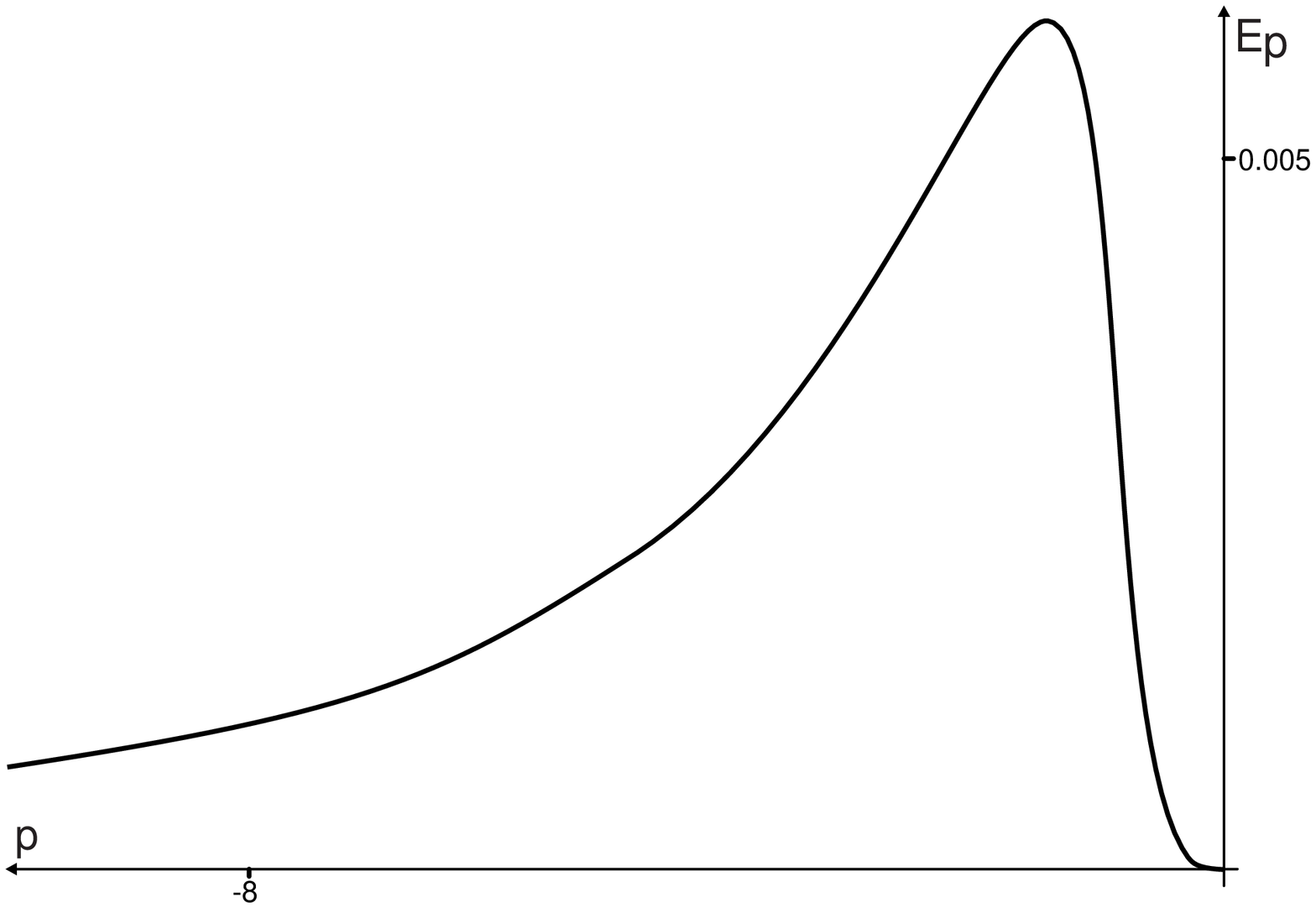}
\caption{Plots of the $p=-1$ and $k=1$ solutions (left), showing the kink (solid line) and anti-kink (dashed line), the corresponding energy density (center, left) and the potential of the Schr\"odinger-like equation (center, right), and the total energy (right) as a function of $p$ for $p$ negative.}
\end{figure}

\subsection{Type-III models}
\label{sec:iii}

An interesting modification of the potential investigated above is given by
\begin{equation}\label{potdsg6n}
{\wt V}(\chi)=\frac1{2p^2}\chi^{2-2p}\,\cos^2(\chi^p)
\end{equation}
which appears with the deformation function $f(\chi)=\sin(\chi^p)$ applied to the $\phi^4$ model. For $p=1$ we have the normal sine-Gordon case. For this new class of models, we have to restrict the parameter $p$ to the interval $(0,1)$, thus making the potentials vanish at the origin, since the choice $p>1$ would lead to a divergence at $\chi=0$.  The minima occur at ${\bar\chi}=0$ and $\pm |(k+1/2)\pi|^{1/p}$, where $k$ is a natural number. The main difference between this class of models and the previous one is the behavior at the central minimum. In the present case, the minimum at ${\bar\chi}=0$ is non perturbative because the second derivative of ${\wt V}$ is not finite at the origin, that is
\be
\left. \frac{d^2{\wt V}(\chi)}{d\chi^2} \right|_{{\bar\chi}=0} \to \infty .
\ee
This feature of the potential (\ref{potdsg6n}) has appeared before in a model introduced in \cite{bmm}. The potential is plotted in Fig.~10 and,
as shown in Ref.~\cite{bmm}, the non perturbative behavior of the symmetric minimum allows the presence of two-kink solutions. These solutions,
which connect the minima $-(\pi/2)^{1/p}$ and $(\pi/2)^{1/p}$, are given by
\be\label{2kinks}
\chi(x)=\pm\; {\rm sign}(x)\left(\arcsin(\tanh(|x|))\right)^{1/p} ,
\ee
and the corresponding energy density is given by
\be\label{2kinksrho}
\rho(x)=\frac{1}{p^2}\arcsin(\tanh(|x|))^{2/p-2}\;\sech^2(x) .
\ee
The two-kink solution and its energy density are also plotted in Fig.~11 for the value $p=1/3$.
The others one-kink solutions are given by
\be\label{1kink}
\chi(x)=\pm\left(k\pi\pm\arcsin(\tanh(x))\right)^{1/p} ,
\ee
and the corresponding energy density is given by
\be\label{1kinksrho}
\rho(x)=\frac{1}{p^2}\left(k\pi\pm\arcsin\left(\tanh(x)\right)\right)^{2/p-2}\;\sech^2(x) ,
\ee
where $k=1,2,3,..$.

\begin{figure}[ht]
\includegraphics[{height=4cm,width=8cm}]{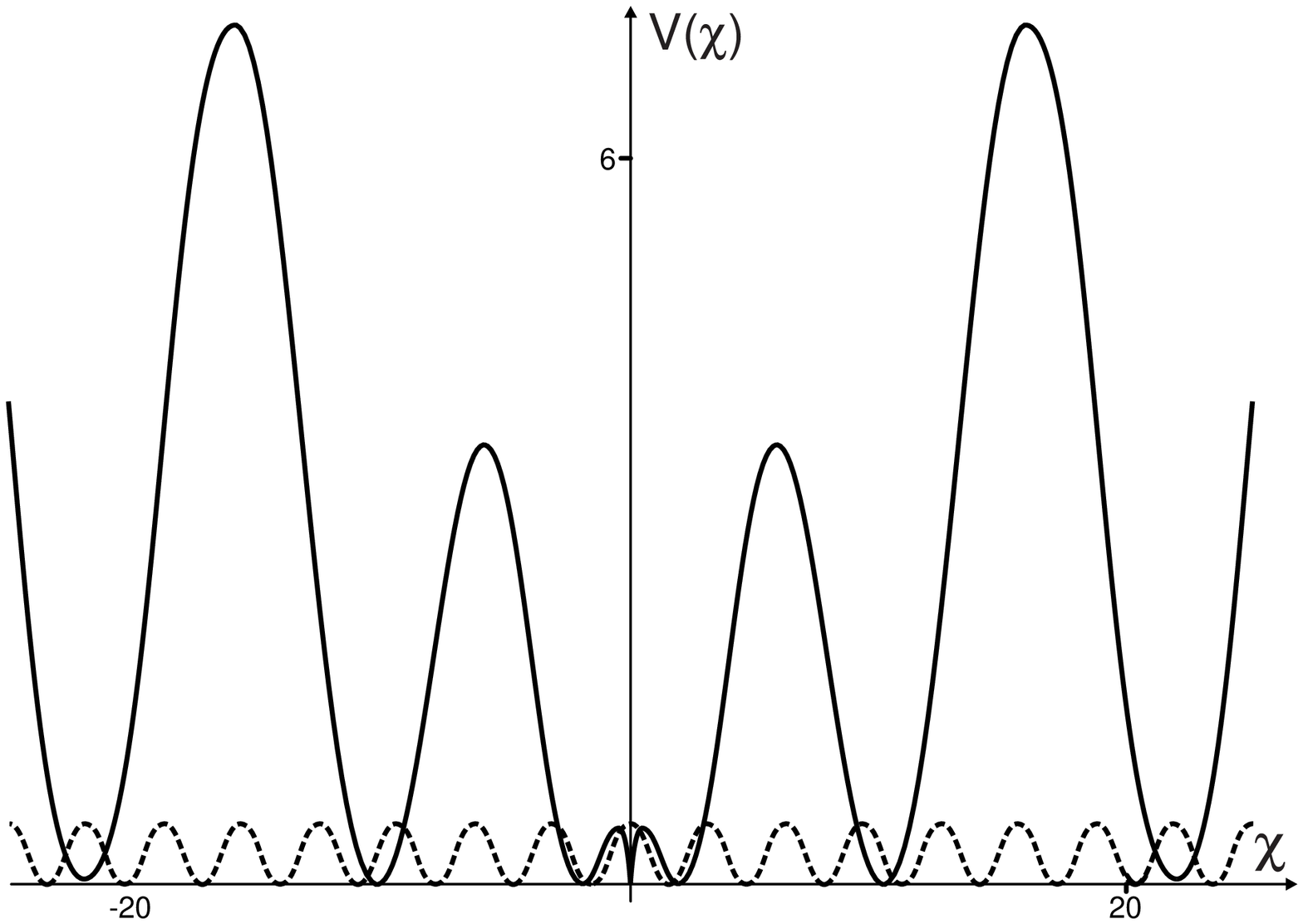}
\hspace{0.5cm}
\includegraphics[{height=4cm,width=8cm}]{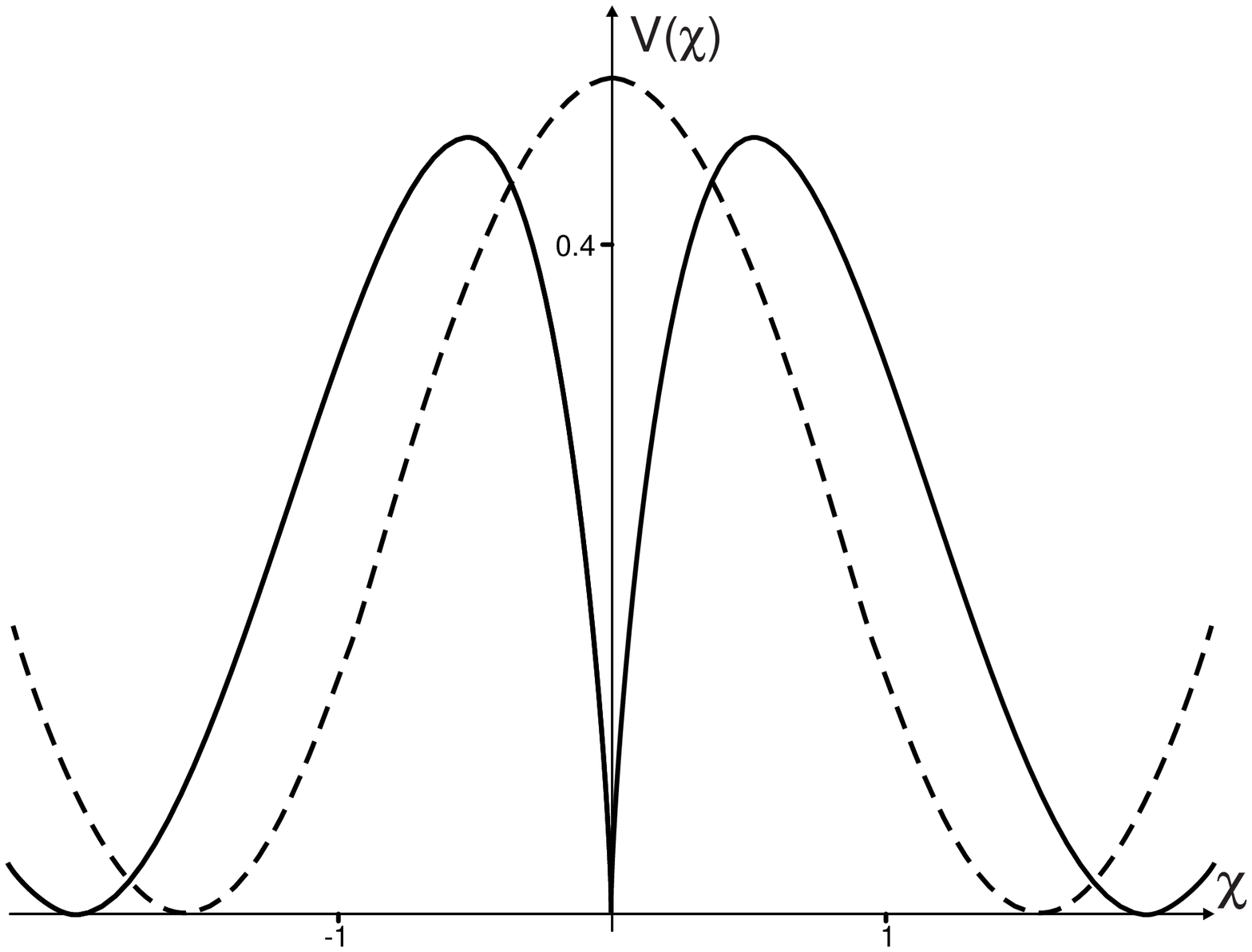}
\caption{The potential (\ref{potdsg6n}) (left) for $p=2/3$ (solid line) and $p=1$ (dashed line), and the behavior near the origin (right), where the symmetric minimum makes the second derivative of the potential divergent.}
\end{figure}
For $0<p\leq 1/2$ the energy of the solutions (\ref{2kinks}) in terms of hypergeometric function is given by
\be
E_p=\frac{2^{2-\frac2p}\;\pi^{\frac2p-1}}{p(2-p)} \mbox{ }_1
F\left(\frac1p-\frac12;\frac12,\frac{p+2}p; -\frac{\pi^{2}}{16} \right).
\ee
The energy of this solution as function of $p$ is presented in fig.~11. The energy diverges for $p\rightarrow0,$ and when $p\rightarrow1$ we have $E_1\rightarrow2$ . Some values of the energy
for $0<p\leq1/2$ are: $E_{1/5}=49.87$, $E_{1/4}=20.61$, $E_{1/3}=8.62$, and $E_{1/2}=3.74$
\begin{figure}[ht]
\includegraphics[{height=4cm,width=5.5cm}]{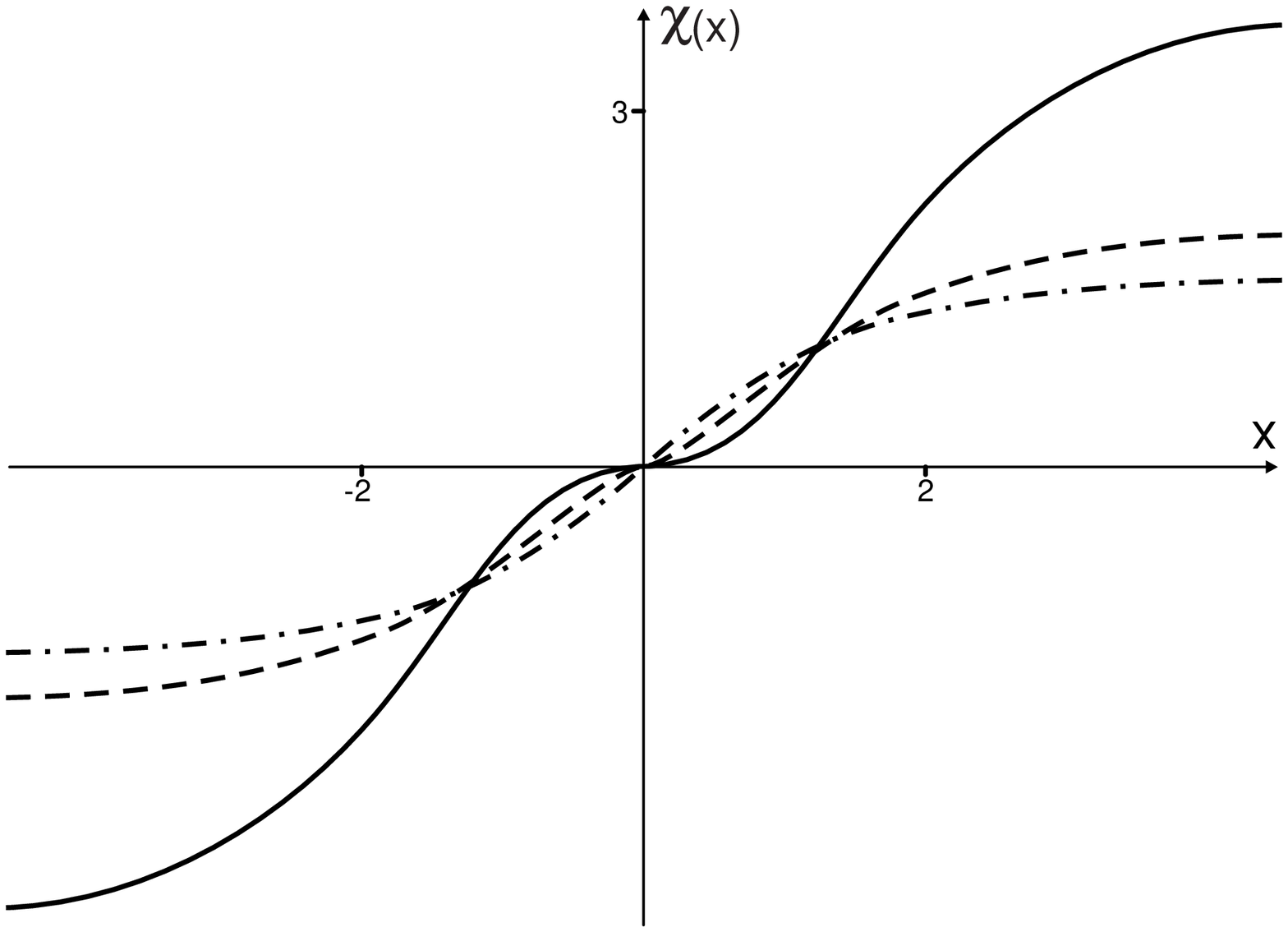}
\includegraphics[{height=4cm,width=5.5cm}]{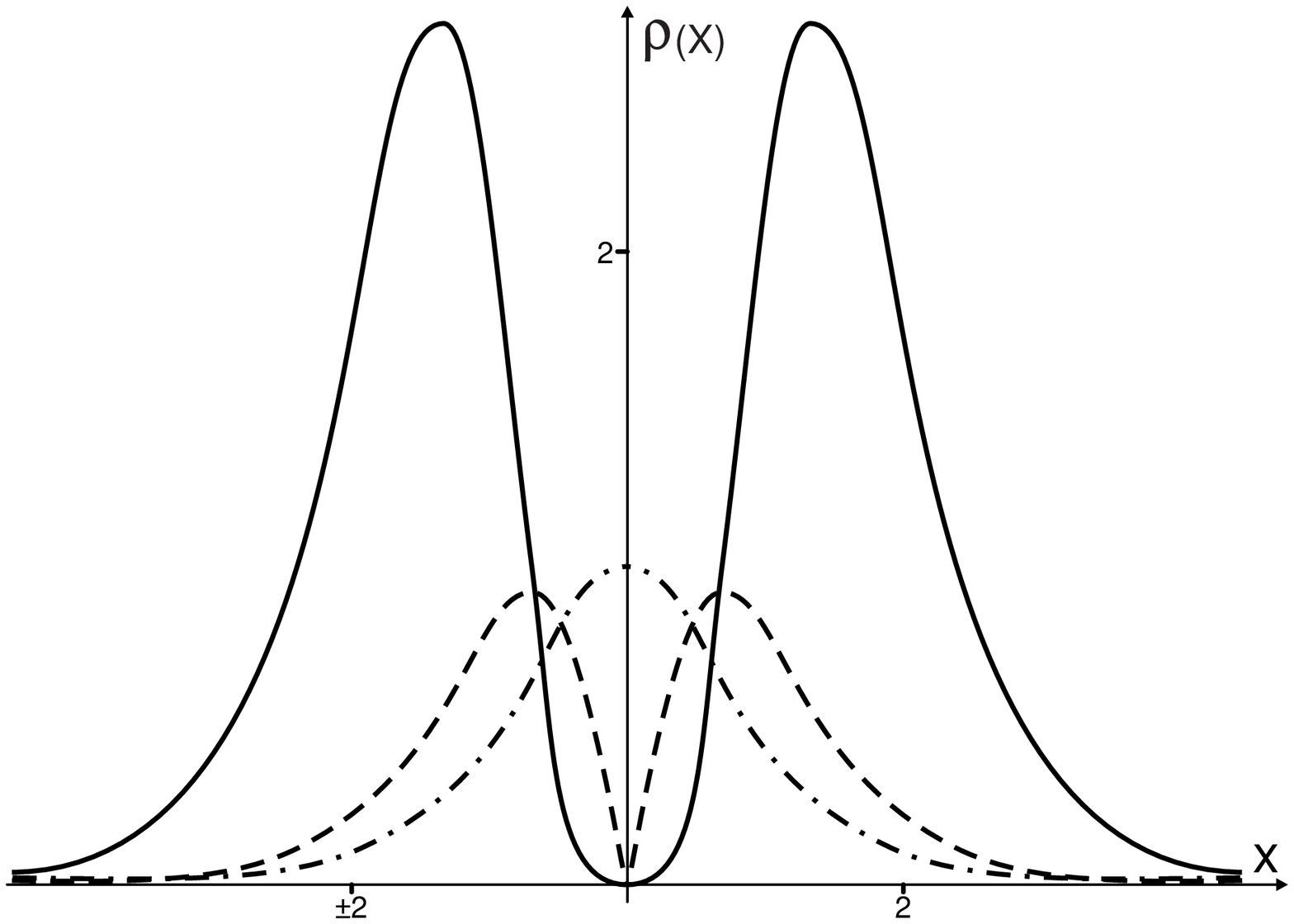}
\includegraphics[{height=4cm,width=5.5cm}]{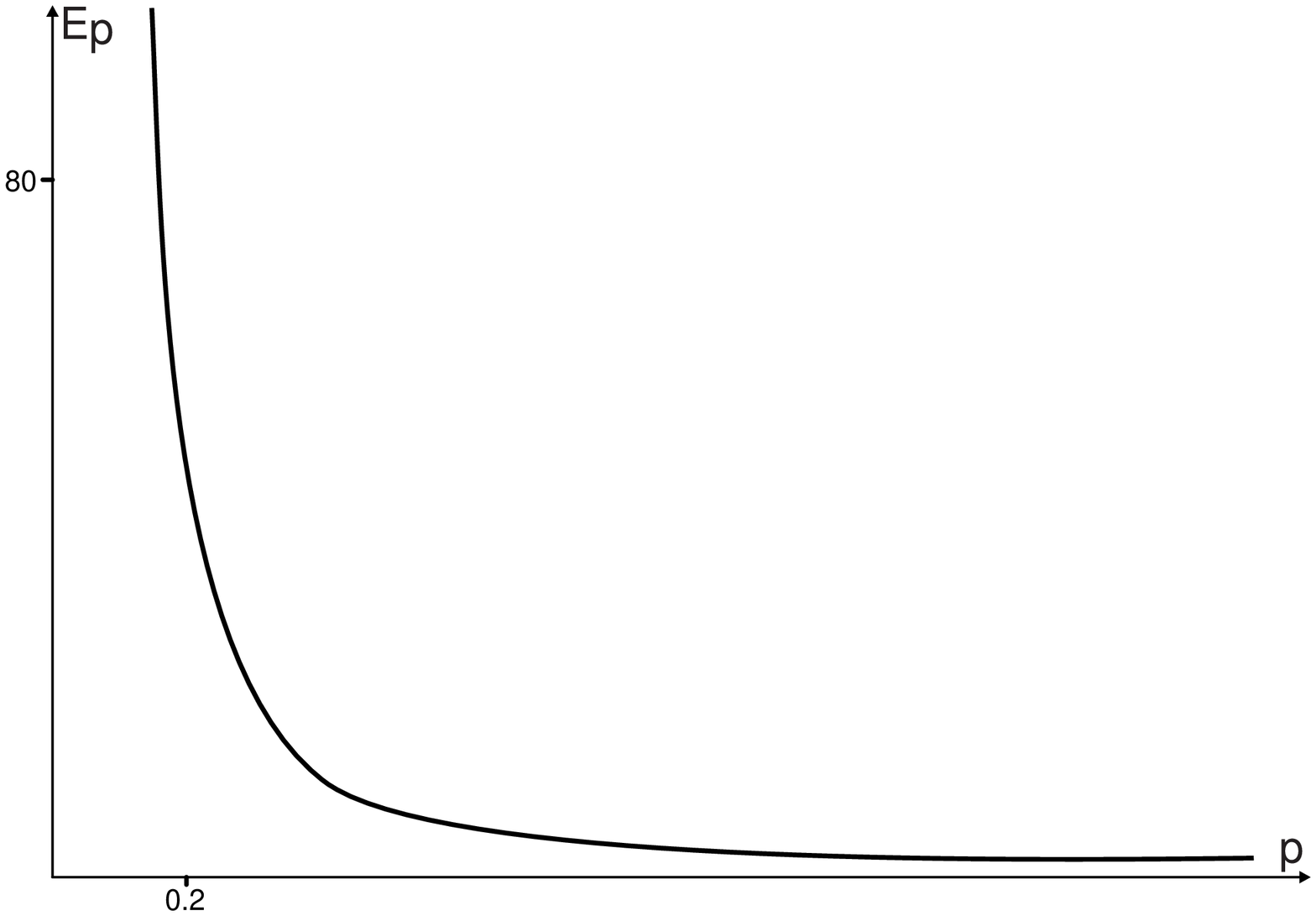}
\caption{The kink-like solutions (\ref{2kinks}) (right) for the cases $p=1/3$ (solid line), $p=2/3$ (dashed line) and $p=1$ (dash-dotted line), the corresponding density energy \eqref{2kinksrho} (center), and the energy of the two-kink solution as a function of $p$ (left).}
\end{figure}

The potential of the Schr\"odinger-like equation are done by the relation \eqref{qm2}, like for Type-II models, but here $a_k(x)=k\pi+{\rm arcsin}\left(\tanh(x)\right)$ and $a_{\bar k}(x)=k\pi-{\rm arcsin}\left(\tanh(x)\right)$. They are plotted in Fig.~12, for $p=1/3$ and $p=1$. We can see that it goes to \eqref{sgq} in the limit $p\to1,$ as expected. 
\begin{figure}[ht]
\includegraphics[{height=4cm,width=6cm}]{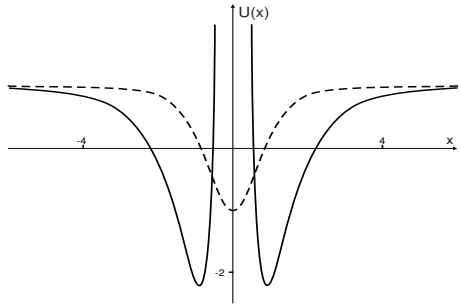}
\caption{The potential of the Schr\"odinger-like equation \eqref{qm2} of two-kink solution for the case $p=1/3$ (solid line) compared with that of one-kink for $p=1$ (dashed line).}
\end{figure}

The two-kink solutions only appear connecting the minima $\pm(\pi/2)^{1/p},$ tunneling through the central symmetric minimum at $\bar\chi=0.$ All the other adjacent minima give rise to standard topological sectors. In those sectors, the potentials of the Schr\"odinger-like equations are very similar to the potentials shown in the right panel in Fig.~7, and can also be approximated by the potential of the Schr\"odinger-like equation which appears from the sine-Gordon model. 

In the present model, we have shown how to find the exotic two-kink behavior first introduced in \cite{bmm} for non-polynomial potentials. This result is also of interest, since it was recently shown how to experimentally construct this type of solution in magnetic systems, with geometry specifically constrained to make them appear as interfaces separating distinct regions of different magnetization \cite{cg}.  

\section{Final comments}
\label{s4}

In this work we have investigated the presence of defect structures in several models of the sine-Gordon type. All the models are obtained as deformations of the $\phi^4$ model, and are constructed with non-polynomials functions. The models investigated in this work are different from the sine-Gordon model, and we distinguish every one with the deformation function used in each case. Specifically, we note that each deformation function is in general defined with a real parameter, and we can use such parameter as a deformation parameter, to control the way the new potential differs from the sine-Gordon model.  
The above identification can be nicely used in applications where the sine-Gordon model is important, since now we can use the deformation parameter to better control the model. In this sense, the models that we have just investigated will certainly enlarge applicability of the sine-Gordon model.

Another interesting issue concerns integrability. Apparently, all the above sine-Gordon-like models seem to loose integrability, and it would be interesting to know how to deal with them at the semiclassical level. We believe that the method developed in Ref.~\cite{g2}, which has already made advances with the double sine-Gordon model, would also work nicely for the models here introduced. Another issue of interest is related with the possible presence of internal modes. This problem appears in the double sine-Gordon model \cite{g1}, and may be investigated considering the model as a non-integrable deformation of the sine-Gordon model. The new models that we have introduced above may engender similar features, and may be studied similarly. These and other related issues are presently under consideration, and we hope to report on them in the near future.    

Among other things, we believe that the deformation procedure is also useful in the sense that it engenders the mapping of solutions of the starting model
into the many solutions of the deformed model in a systematic way. Such a feature has already appeared in the polynomial potentials introduced in \cite{dd4}, which were investigated in \cite{g22} to unveil the presence of neutral states very nicely. In this sense, we hope that the models we have built may be of good use both theoretically and phenomenologically.   

The authors would like to thank CAPES, CNPq, and PRONEX/CNPq/FAPESQ for financial support.


\end{document}